\documentstyle[epsf]{mn}
\input epsf.sty
\def\hmpc{~h$^{-1}$ Mpc~}

\title[A study of the core of the Shapley Concentration V]
 { A study of the core of the Shapley Concentration: \\
 V. The A3528 complex: a young merger event?
\thanks{based on observations collected at the European Southern
Observatory, La Silla, Chile.} }
\author[S.Bardelli, E.Zucca \& A.Baldi]
{
Sandro Bardelli$^{1}$,
Elena Zucca$^{1}$
\&
Alessandro Baldi$^{2}$
\\ $^1$ Osservatorio Astronomico di Bologna, 
via Ranzani 1, I--40127 Bologna, Italy
\\$^2$ Dipartimento di Astronomia, Universit\`a degli Studi di Bologna,
via Ranzani 1, I--40127 Bologna, Italy
\\ E-mail: bardelli@bo.astro.it, zucca@bo.astro.it
}
\date{Received 00 - 00 - 0000; accepted 00 - 00 - 0000}
\begin{document}
\maketitle
\begin{abstract}

We present the results of a redshift survey of galaxies in the A3528 complex,
a chain of interacting clusters in the core of the Shapley Concentration.
This complex is characterized in the X-ray band by two pairs of roughly 
similar interacting clumps: one pair has been resolved as two optical Abell
clusters (A3530 and A3532), while the two components of the other pair
are both associated to A3528. The optical data show that the distance between 
the centers of A3530 and A3532 is smaller than their Abell radii, an indication 
of the existence at least of tidal interactions, and that the contours of 
galaxies in A3528 appear to be elongated in the North--South direction, 
pointing towards the A3530--A3532 pair.

From our survey we obtained $\sim 600$ new radial velocity determinations: 
using this sample, we studied the dynamics of the four Abell clusters in this
region (A3528, A3530, A3532 and A3535) and derived their mean velocities and
velocity dispersions. Moreover we performed a substructure analysis, both
bi--dimensional and three--dimensional, of the whole complex. 

All the characteristics of this structure seem to point toward a
merging scenario in an early stage, whose effects on the galaxy population 
and on the cluster dynamics are not yet evident, contrary to what happens
in the nearby A3558 complex, where the merging events are in a rather advanced 
stage and were already able to induce modifications in the galaxy population.

\end{abstract}


%-------------------------------------------------------------------------------
%
\section{Introduction}
Clusters of galaxies are thought to form by accretion of subunits in 
a hierarchical bottom-up scenario (see f.i. Colberg et al. 1998): the fact
that substructures are observed in a high fraction of clusters (see f.i. Bird 
1994) indicates that this accretion phenomenon is still active.  
\\
Cluster mergings are among the most energetic phenomena in the Universe, 
leading to a release of $\sim 10^{50-60}$ ergs in a time scale of 
the order of Gyrs. 
The astrophysical consequences of this energy output are expected to be both 
thermal (shocks, changes in the temperature and gas distribution profiles
of the Intra Cluster Medium) and non--thermal (appearance of radio sources like
Halos or Relics, hard X-ray tails). Such effects are visible at different
wavelengths, but while the merging event is rather well studied in the X-ray 
band, the studies of the velocity distribution and morphological modification 
of the galaxy population suffered until recently by the limited number of 
available spectra per cluster.
A significant improvement in this field will be achieved
by the new generation of multi-fiber spectrographs, which allow to obtain
hundreds of spectra simultaneously. However, even with the ``old" 
multi-fiber spectrographs like OPTOPUS and MEFOS it has been possible to
map major merger events like the A3558 complex in the core of the Shapley
Concentration (Bardelli et al. 1994, 1998a, 1998b, 2000), as part of a large,
multiwavelength project.    

%-------------------------------------------------------------------------------
% FIGURE 1 
\begin{figure*}
\epsfxsize=\hsize 
\epsfbox{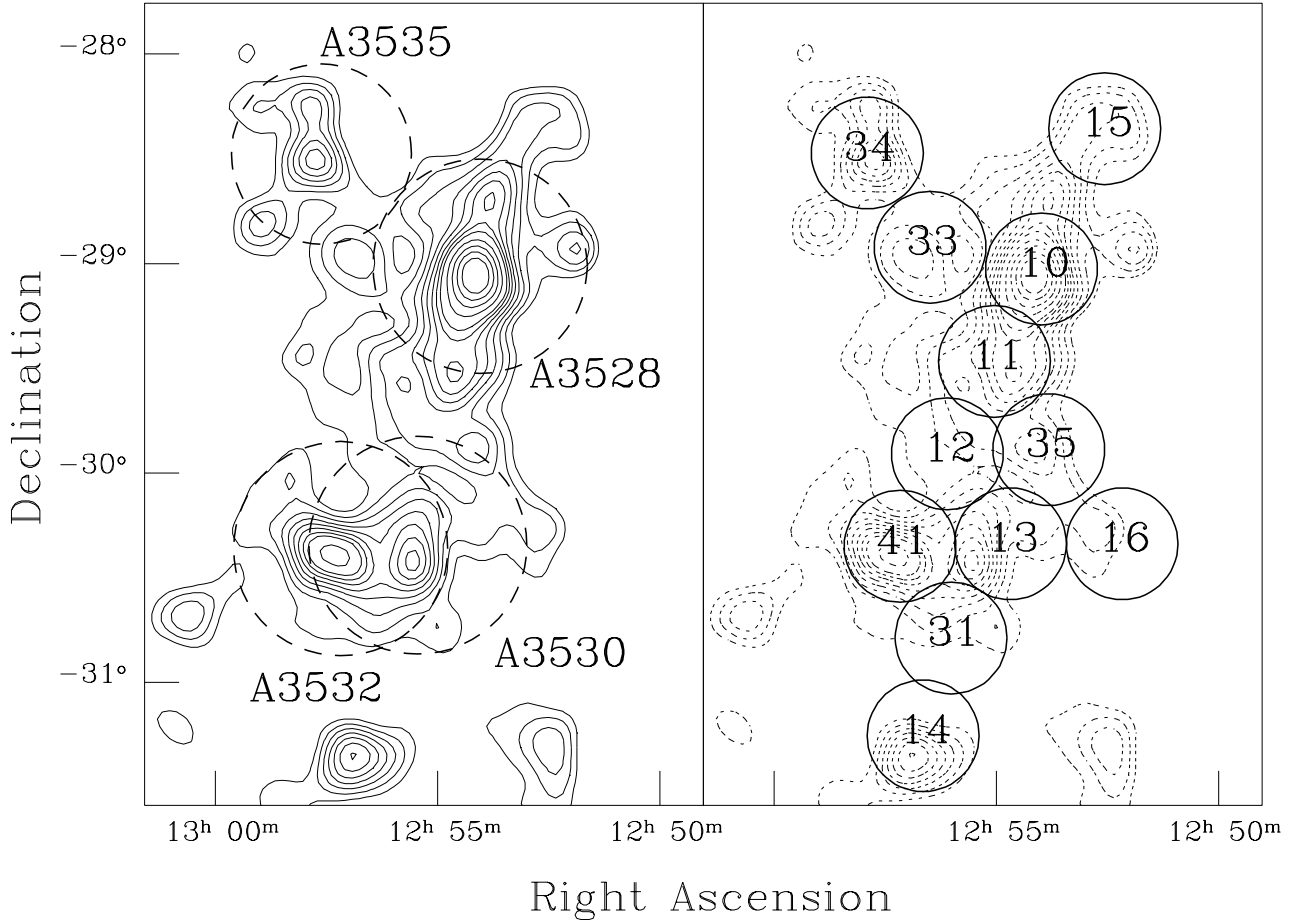}
\caption[]{Left: Isodensity contours of galaxies in
an area of $\sim 2^o.7 \times 3^o.8$ around the A3528 cluster complex.
The figure refers to galaxies in the magnitude range $17-19.5$ and the data
have been binned in $2\times 2$ arcmin$^2$ bins and smoothed with a Gaussian
with a FWHM of $6$ arcmin. For the four ACO clusters in the region circles of 
one Abell radius have been drawn (dashed curves). Right: The same as left panel
with superimposed the observed OPTOPUS fields. }
\label{fig:fields}
\end{figure*}
%-------------------------------------------------------------------------------

The Shapley Concentration is a unique laboratory where to study the cluster 
merging because, given the high local overdensity (of the order of
$\sim 10$ on scales of $10$ \hmpc, Bardelli et al. 2000), the induced 
peculiar velocities increase the ``cross section" for cluster-cluster 
collisions. 
In particular, two clusters complexes are found at the center of the 
supercluster (Zucca et al. 1994), dominated by A3558 and A3528 respectively: 
these structures, whose spatial scales are of the order of $\sim 5$ \hmpc 
(Bardelli et al. 2000), are formed by strongly interacting clusters. 
The complex dominated by A3558 
has been extensively studied in optical (Bardelli et al. 1994, 1998a, 
1998b), radio (Venturi et al. 1997, 1998a, 2000) and X-ray (Bardelli et al.
1996, Ettori et al. 2000) bands, revealing its complex dynamical activity.
We suggested that this structure could be the remnant of a cluster-cluster 
collision seen just after the first core-core encounter.

We are carrying on an analogous multiwavelength study on the complex formed by 
the ACO (Abell, Corwin \& Olowin, 1989) 
clusters A3528, A3530, A3532 and A3535, which appears very similar to 
the A3558 complex. These two complexes are separated by $\sim 19$ \hmpc and 
are connected by a bridge of galaxies (Bardelli et al. 2000). 
In this paper we present the results of a dynamical and substructure 
analysis of the clusters in this complex, based on $\sim 600$ new velocity
determinations.

The plan of the paper is the following: in Sect. 2 we present the
characteristics of the A3528 complex and in Sect. 3 we describe our galaxy 
sample. In Sect. 4 we derive the dynamical parameters of the clusters
and in Sect. 5 we analyze their substructures.
Finally the results are discussed in Sect. 6.

%-------------------------------------------------------------------------------
%
\section{The A3528 cluster complex}

Fig. \ref{fig:fields} shows the galaxy isodensity contours in the sky region
covering the A3528 complex.
This figure shows two interesting features: a) the distance between 
the centers of A3530 and A3532 is smaller than their Abell radii, an 
indication of the existence at least of tidal interactions; b) the contours
of galaxies in A3528 appear to be elongated in the North--South direction, 
pointing towards the A3530--A3532 pair.
The complex extends for $\sim 2.8$ degrees in the North--South and $\sim 1.7$ 
degrees in the East--West directions, corresponding to linear sizes of
$\sim 8$ \hmpc and $\sim 5$ \hmpc, respectively, at the mean redshift of
the structure (see below).

The cluster A3528 
($\alpha_{2000}=12^h 54^m 34^s$, $\delta_{2000}=-29^o 08' 30''$)
is of richness class 1: all the other clusters of the complex
are of richness class 0. Quintana et al. (1995) found
a mean velocity of $<v>=15631\pm 148$ km/s and a velocity dispersion of
$\sigma=864^{+119}_{-85}$ km/s determined with 39 galaxies.
The ENACS project estimated for this cluster $<v>=15780$ km/s and 
$\sigma=969$ km/s, using 28 galaxies (Mazure et al. 1996).
\\
From a ROSAT PSPC observation, Schindler (1996) found that this
cluster is actually double, formed by two subcomponents 
(dubbed A3528N and A3528S) showing an indication
of interaction. In fact, the gas seems to be hotter in the region between 
the two subclusters. Henricksen \& Jones (1996) estimated 
an hot gas temperature of $2.7\pm 0.8$ and $2.9\pm 0.9$ keV
for A3528N and A3528S, respectively. The core radii are $88\pm 10$
and $58 \pm 5$ h$^{-1}$ kpc, while the total masses are $3.1$ and $3.9 \times
10^{13}$ h$^{-1}$ M$_{\odot}$.
Radio observations of this cluster (Schindler 1996; Reid, Hunstead \& Pierre
1998; Venturi et al. 1998b) show that in this cluster and in 
its proximities a large number of extended radio galaxies is located,
sign of dynamical activity.

The cluster A3530 
($\alpha_{2000}=12^h 55^m 31^s$, $\delta_{2000}=-30^o 19' 53''$)
has an estimated mean velocity of $<v>=15850\pm 165$ km/s and a velocity 
dispersion of $\sigma=643^{+146}_{-88}$ km/s, based on 18 galaxies 
(Quintana et al. 1995). 
White, Jones \& Forman (1997) reported for this cluster a temperature for the 
hot gas of $3.2$ keV and Ettori, Fabian \& White (1997) estimated a total mass 
of $1.13 \times 10^{14}$ h$^{-1}$ M$_{\odot}$. 
   
The estimated mean velocity and dispersion of the cluster A3532 
($\alpha_{2000}=12^h 57^m 22^s$, $\delta_{2000}=-30^o 22' 03''$) are
$<v>=16085\pm 102$ km/s and $\sigma=594^{+82}_{-58}$ km/s, respectively,
based on 39 galaxies (Quintana et al. 1995). 
White et al. (1997) reported a temperature of $4.4 \pm 1.5$ keV and Ettori et 
al. (1997) a total mass of $1.7 \times 10^{14}$ h$^{-1}$ M$_{\odot}$.

The cluster A3535 
($\alpha_{2000}=12^h 57^m 48^s$, $\delta_{2000}=-28^o 29' 18''$), 
is the least studied of the complex. The only data available in the
literature concern its redshift: on the basis of 4 radial velocities
Vettolani et al. (1990) estimated a mean velocity of 19770 km/s. 

%-------------------------------------------------------------------------------
%
\section{The galaxy sample}

Fig. \ref{fig:fields} shows the isodensity contours for the objects in the 
magnitude bin $17\le b_J \le 19.5$ from the COSMOS/UKSTJ galaxy catalogue
(Yentis et al. 1992), in a region of $\sim 2^o.7 \times 3^o.8$,
corresponding to $12^h 51^m < \alpha_{2000} <13^h 15^m $ and 
$-31^o 36' < \delta_{2000} < -28^o 00'$.
This region is part of the plate 443, nearby the border with the plate 442,
which is located on the West. 
The data have been binned in cells of $2\times 2 $ arcmin$^2$
and then smoothed with a Gaussian with 6 arcmin of FWHM. In the left panel,
circles of one Abell radius have been drawn on the four Abell clusters.
In the right panel, the positions of the observed OPTOPUS fields are
shown. In Table \ref{tab:fields}
the coordinates and the observation dates of the fields are reported.

%-------------------------------------------------------------------------------
% TABLE 1.
\begin{table}
\caption[]{ Centers of the observed OPTOPUS fields }
\begin{flushleft}
\begin{tabular}{rrrr}
\hline\noalign{\smallskip}
Field & $\alpha$ (2000) & $\delta$ (2000)& Obs. Date   \\
\noalign{\smallskip}
\hline\noalign{\smallskip}
  10  & 12 54 18 & -29 01 00 & ~~~~8/9-Mar-1991  \\
  11  & 12 55 15 & -29 28 42 & 24/27-Feb-1993 \\
  12  & 12 56 16 & -29 55 12 & 24-Feb-1993    \\
  13  & 12 54 20 & -30 21 00 & 22/27-Feb-1993 \\
  14  & 12 56 45 & -31 16 00 & 26-Feb-1993    \\
  15  & 12 52 54 & -28 22 88 & 27-Feb-1993    \\
  16  & 12 52 54 & -30 21 00 & 26-Feb-1993    \\
  31  & 12 56 09 & -30 48 00 & 27-Feb-1993    \\
  33  & 12 56 42 & -28 56 00 & 25-Feb-1993    \\
  34  & 12 58 05 & -28 29 00 & 23/26-Feb-1993 \\
  35  & 12 54 02 & -29 54 00 & 25-Feb-1993    \\
  41  & 12 57 00 & -30 21 42 & 26-Feb-1993    \\
 \noalign{\smallskip}
\hline
\end{tabular}
\end{flushleft}
\label{tab:fields}
\end{table}
%-------------------------------------------------------------------------------

%-------------------------------------------------------------------------------
% FIGURE 2 
\begin{figure}
\epsfxsize=\hsize 
\epsfbox{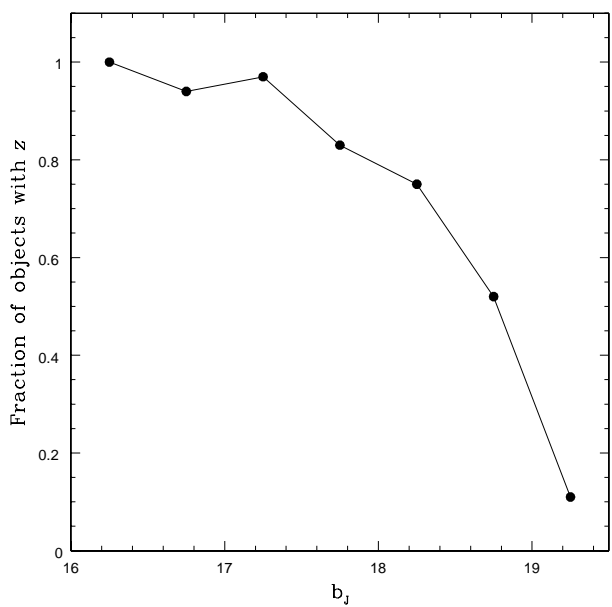}
\caption[]{ Redshift completeness of the spectroscopic sample as a function of 
magnitude. Note that the first bin refers to all galaxies with $b_J<16.5$. }
\label{fig:completezza}
\end{figure}
%-------------------------------------------------------------------------------

%-------------------------------------------------------------------------------
%
\subsection{Observations and data reduction}

The spectroscopic observations were performed at the 3.6m ESO telescope 
at La Silla, equipped with the OPTOPUS multifiber spectrograph (Lund 1986), on 
the nights of 8 and 9 March 1991 (for the field \# 10) and
from  23 to 27 February 1993. Fields \# 10, \# 11, \# 13 and \# 34
have been observed twice given the high density of galaxies 
and/or lack of velocity determination in the literature. 

The OPTOPUS multifiber spectrograph uses bundles of 50 optical 
fibers, which can be set within the field of the Cassegrain focal plane 
of the telescope; this field has a diameter of $32$ arcmin, and each fiber
has a projected size on the sky of $\sim 2.5$ arcsec.
We used the ESO grating $\#\ 15$ 
($300$ lines/mm and blaze angle of $4^o 18'$) allowing a dispersion of 
$174$ \AA /mm (corresponding to a resolution of $\sim 12$ \AA) in the 
wavelength range from $3700$ to $6024$ \AA. 
The detectors were Tektronic $512\times 512$ CB CCDs (ESO $\# 16$ for 1991 run;
ESO $\# 32$ for 1993 run) with a 
pixel size of $27\ \mu$m, corresponding to $4.5$ \AA, i.e. a velocity bin of 
$\sim 270$ km/s at $5000$ \AA. 
Detector $\# 32$ has a particularly good responsive quantum function in the 
blue ($\sim 70\%$ at $4000$ \AA), where there are the calcium and [OII] lines.

Following Wyse \& Gilmore (1992), we dedicated 5 fibers to sky measurements, 
remaining with 45 fibers available for the objects.
The observing time for each field was one hour, split into two half--hour
exposures in order to minimize the effects due to the ``cosmic'' hits. The 
observing sequence was: a 30 seconds exposure of a quartz--halogen white lamp,
a 180 seconds exposure of the Phillips Helium and Neon arcs, then the first and 
the second field exposures, and again the arcs and the white lamp. 

%-------------------------------------------------------------------------------
% FIGURE 3 
\begin{figure}
\epsfxsize=\hsize 
\epsfbox{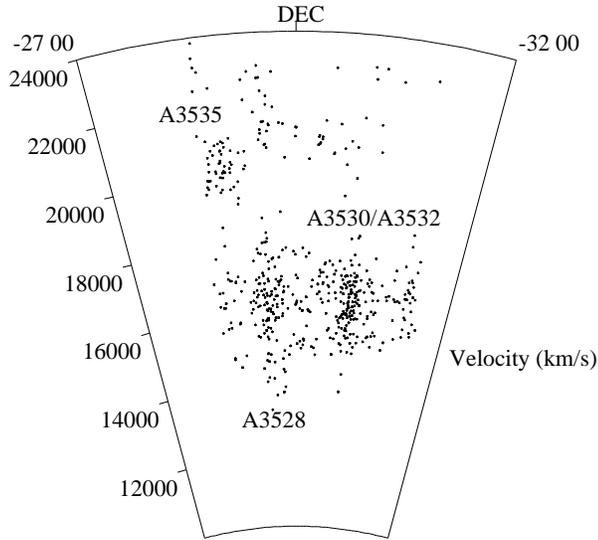}
\caption[]{ Wedge diagram for galaxies in the
velocity range $11000-24000$ km/s. The diagram is compressed in right 
ascension and displayed in declination, in order to better visualize the
structure. }
\label{fig:wedge}
\end{figure}
%-------------------------------------------------------------------------------

The reduction steps are described in Bardelli et al. (1994). 
However, it could be important to
stress that we normalized the fiber transmission dividing each spectrum by
the continuum--subtracted flux of the sky emission line [OI]$\lambda$ 5577. 
This procedure assumes that the sky emission does not change significantly
on angular scales of the order of $\sim 30$ arcmin (Wyse \& Gilmore 1992). 
 
We obtained estimates of the radial velocity of galaxies using the 
cross--correlation method as implemented in the IRAF 
\footnote{IRAF is distributed by 
KPNO, NOAO, operated by the AURA, Inc., for the National Science Foundation.}
task RVSAO (XCSAO, Kurtz et al. 1992). 
The galaxy spectra were cross--correlated with those of $8$ stellar templates
observed with the same instrumental set up. 
The adopted velocity for each galaxy is the value given by the template
which gives the minimum cross--correlation error. For spectra with strong
emission lines we measured an ``emission velocity" using 
the EMSAO program in the same IRAF task RVSAO.

%-------------------------------------------------------------------------------
% FIGURE 4
\begin{figure}
\epsfxsize=\hsize 
\epsfbox{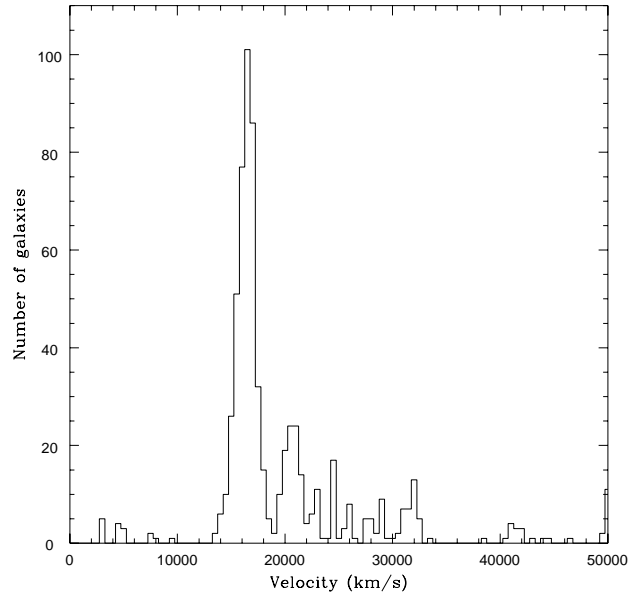}
\caption[]{Velocity histogram of the galaxies in the range $0-50000$ km/s: 
the main peak at $\sim 16000$ km/s corresponds to A3528, A3530
and A3532, while the peak at $\sim 20000$ km/s corresponds to A3535. }
\label{fig:allhisto}
\end{figure}
%-------------------------------------------------------------------------------

%-------------------------------------------------------------------------------
% FIGURE 5
\begin{figure*}
\centering
\leavevmode
\epsfxsize=0.49\hsize \epsfbox{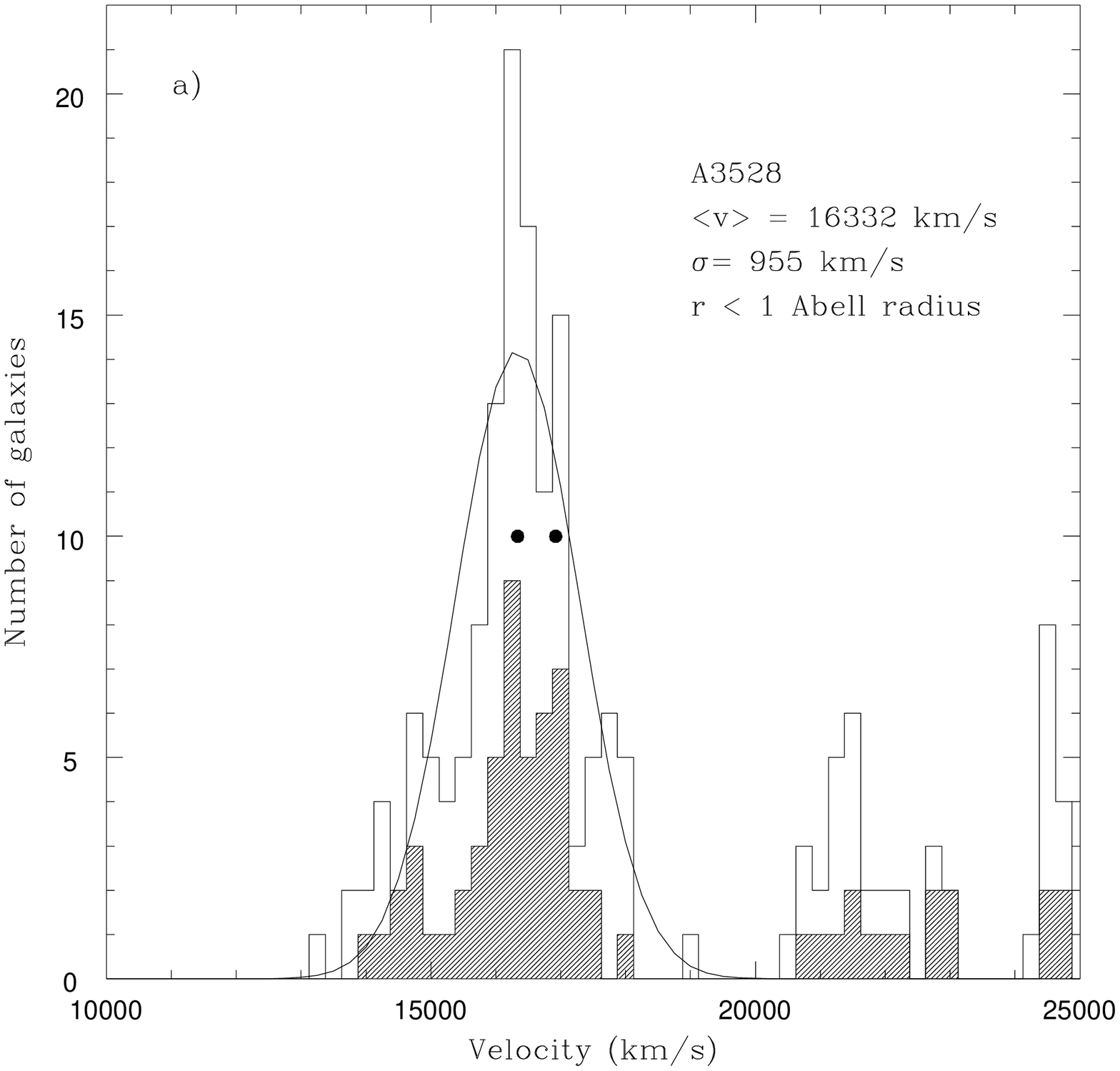} \hfil
\epsfxsize=0.49\hsize \epsfbox{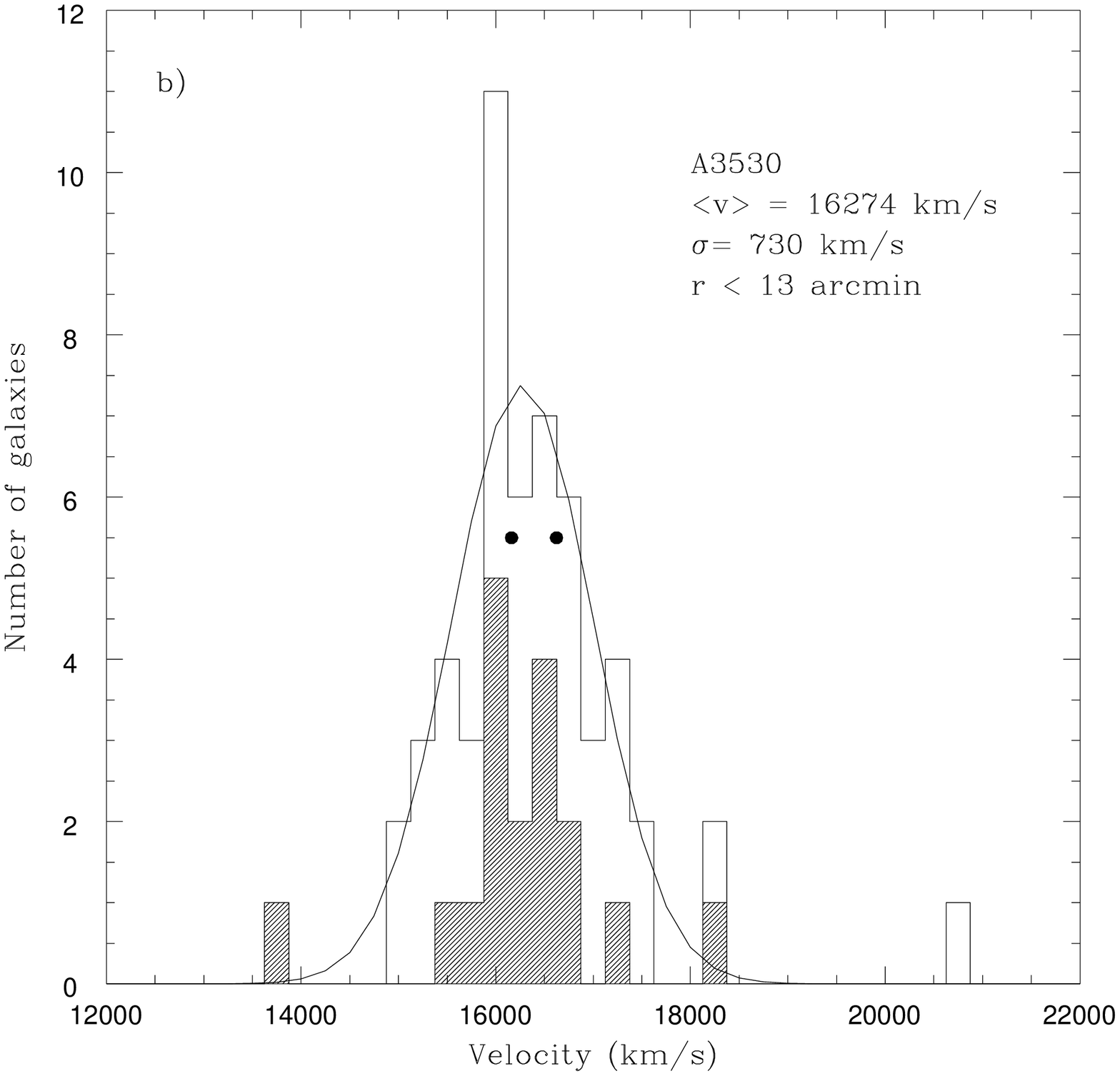} \hfil
\centering
\leavevmode
\epsfxsize=0.49\hsize \epsfbox{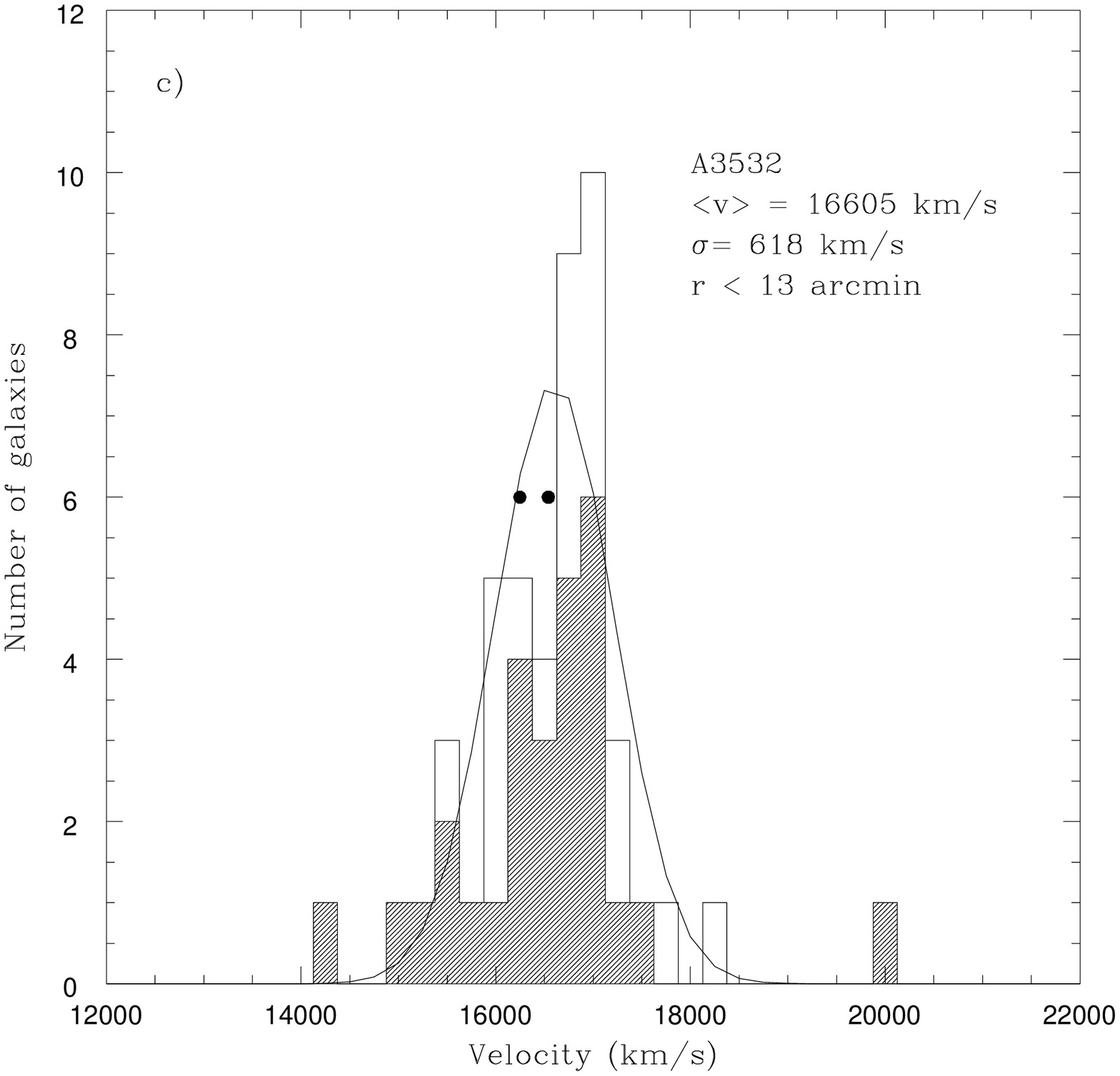} \hfil
\epsfxsize=0.49\hsize \epsfbox{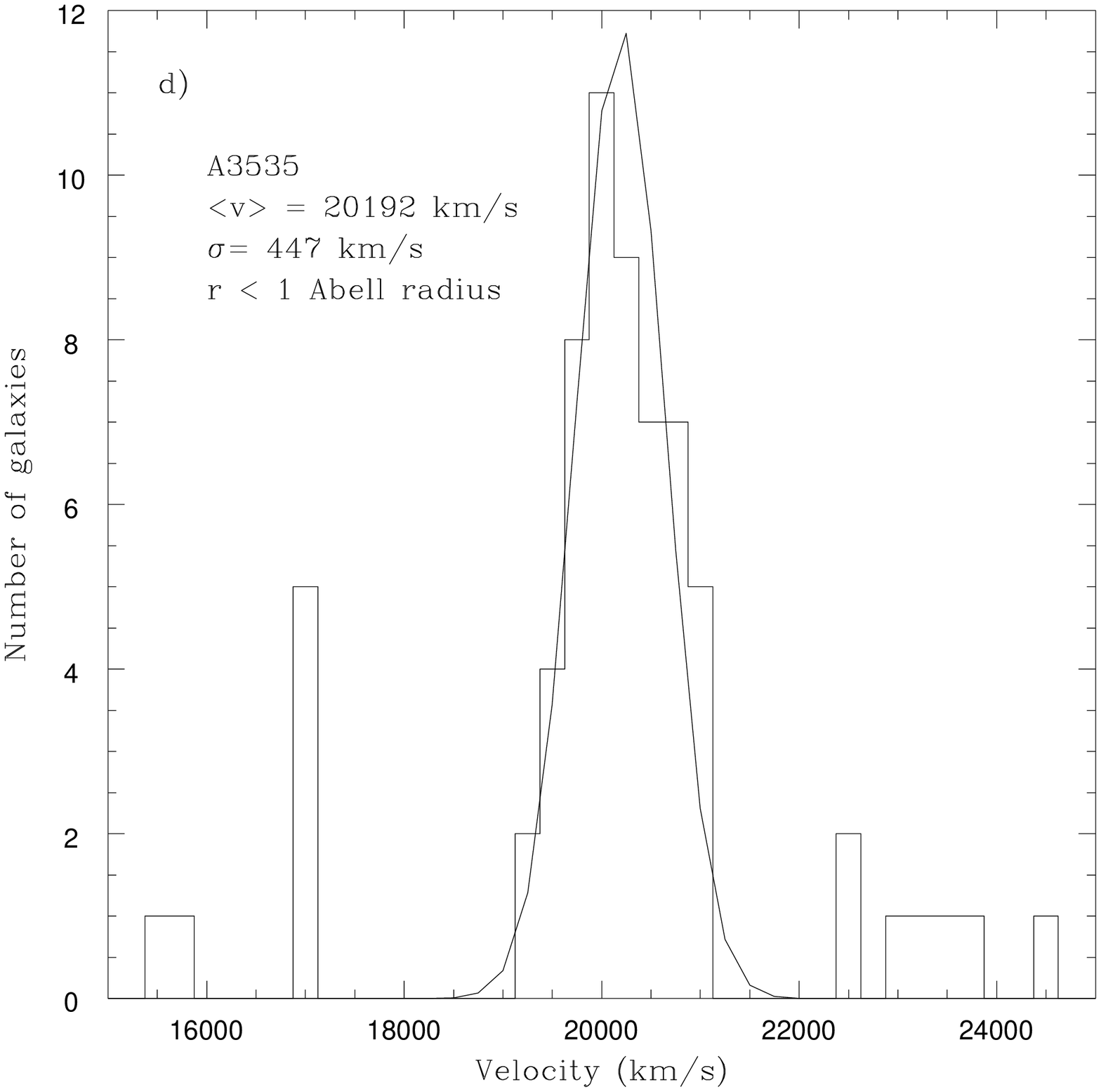} \hfil
\caption[]{Velocity histograms for the four ACO clusters in the complex:
a) A3528; b) A3530; c)A3532; d) A3535. A Gaussian with mean value $<v>$ and
dispersion $\sigma$ has been superimposed on each distribution: the 
parameters for each cluster are reported in the labels. The shadowed histograms
in the first three panels correspond to galaxies with $b_J < 18$. }  
\label{fig:clhisto}
\end{figure*}
%-------------------------------------------------------------------------------

%-------------------------------------------------------------------------------
%
\subsection{The new redshift data}

From the total number of spectra ($692$), it was possible to obtain $662$
velocity estimates: $81$ objects turned out to be stars ($\sim 12\%$ of the
reliable spectra), leaving us with $581$ new galaxy redshifts. 

In Table 2 we list the objects with velocity determination. Columns (2), 
(3) and (4) give the right ascension (2000), 
the declination (2000) and the $b_J$ apparent magnitude, respectively, as 
reported in the COSMOS catalogue. 
Columns (4) and (5) give the heliocentric velocity ($v=cz$) 
and the internal error (in km/s). The external error can be derived by 
multiplying the error in
column (5) by a factor of the order of $1.6-1.9$: the lower value is
obtained by comparing repeated observations of the same galaxies
(Malumuth et al. 1992), while the higher one takes into account also the 
different reduction techniques (see the discussion in Bardelli et al. 1994). 
Finally, the code ``emiss" in column (6) denotes the velocities determined from 
emission lines. 

The galaxies whose spectrum presents detectable emission lines 
(mainly [OII]$\lambda$ 3727, H$\beta$ $\lambda$ 4860, [OIII]$\lambda$ 4959 and
[OIII]$\lambda$ 5007) are $133$, corresponding 
to a percentage of $22\%$, significantly higher than the average value 
of $16\%$ reported by Biviano et al. (1997) for the cluster galaxy 
population. 

From the literature data, we added 30 galaxies from the ENACS survey 
(Katgert et al. 1998), 17 from Quintana et al. (1995) and 43 from the 
compilation of Fairall \& Jones (1991): in case of multiple values for
the same galaxy, we adopted the most recent estimate. The final sample 
contains a total of 671 velocities. 

In Fig. \ref{fig:completezza} we plot the redshift completeness 
within the OPTOPUS fields as a function of magnitude. The first bin
includes all galaxies with $b_J<16.5$. The differential completeness is
higher that $70\%$ for objects with $b_J< 18.5$: at this limit the 
integrated completeness is $\sim 83 \%$. 

In Fig. \ref{fig:wedge}, the wedge diagram of the galaxies of our
sample in the velocity range $11000-24000$ km/s is reported. In this 
plot the right ascension coordinate has been compressed and the plotted
coordinate is declination, in order to better visualize the structure,  
which is elongated in the North--South direction. Note that, in this
representation, A3530 and A3532 appear compressed together. 
From this figure it is clear the presence of
the ``finger-of-God" of the clusters and the physical connection
between them. Note also the region without galaxies in the foreground of
the structure. 

In Fig. \ref{fig:allhisto} the velocity histogram of our sample is shown: 
the main peak at $\sim 16000$ km/s corresponds to A3528, A3530
and A3532, while the peak at $\sim 20000$ km/s corresponds to A3535. 

These data were used in Bardelli et al. (2000) to estimate the density excess
of the A3528 complex with respect to a uniform distribution. We found 
$ \displaystyle{N \over {\bar N}} = 21.4 \pm 1.2$ on a scale of 5.3 \hmpc, 
which corresponds
to a mass of $3.6\times 10^{15}$ $\Omega_o$ h$^{-1}$ M$_{\odot}$, under the
hypothesis that light traces mass. 
Analyzing the galaxy density profile, we found a significant void in the
foreground (at $\sim 10000$ km/s);  
in the background of the structure we detected a void at a mean
velocity of $\sim 35000$ km/s and a significant overdensity at 
$\sim 30000$ km/s which is part of a larger structure (see the discussion
in Bardelli et al. 2000). 

%-------------------------------------------------------------------------------
%
\section{Dynamical parameters of the clusters}

The basic parameters which describe the dynamics of a cluster are the
mean velocity $<v>$ and the velocity dispersion $\sigma$. In order to
estimate these quantities we used the biweight location and scale
estimators (Beers et al. 1990). The velocity range, in which the cluster 
members lie, is found by assuming that the velocity distribution is 
Gaussian and the procedure used for rejecting interlopers is described in 
Bardelli et al. (1994).   

The parameters obtained for the clusters in the complex are reported in
Table \ref{tab:dynamics}: column (1) and (2) give the cluster name and
the number of galaxies used to estimate the dynamical parameters, column (4)
and (5) show the mean velocity and the velocity dispersion (with their
$1\sigma$ errors).
For A3528 and A3535 we considered galaxies inside one Abell radius from the
cluster center (i.e. $31'$ and $25'$ respectively); for A3530 and A3532, given 
the fact that their Abell circles overlap each other, we restricted our 
analysis to galaxies inside a circle of 13 arcmin radius.
In Fig. \ref{fig:clhisto} the velocity histograms for each cluster are shown,
with superimposed a Gaussian with mean value $<v>$ and dispersion $\sigma$;
the shaded histograms correspond to galaxies with $b_J < 18$.   

Comparing our parameters reported in Table \ref{tab:dynamics} with the
literature values (see Sect. 2), we find significant differences.
The velocity distribution of galaxies in A3528, although globally consistent
with a single Gaussian, appears characterized by several peaks. In particular,
considering the distribution of bright galaxies ($b_J < 18$, shadowed 
histogram in Fig. \ref{fig:clhisto}a) the presence of a bump at lower  
velocity ($\sim 14500$ km/s) is clear. 
Given the fact that the ENACS and the Quintana et al.
(1995) estimates are based on galaxies brighter than $\sim 18$, this tail
can explain their lower values for $<v>$. 
An analogous justification can explain the differences for A3530 and A3532,
in addition to the fact that for these two clusters we restricted the
analysis to the very central regions. Finally, for A3535 the difference with
the literature value is justified by the fact that the Vettolani et al. (1990)
estimate was based only on 4 galaxies.

Another interesting characteristics of this complex is that all clusters  
appear dominated by a pair of bright ellipticals, instead of a single 
dominant galaxy.
In particular, A3530 and A3532 are both dominated by a dumb-bell system
(Gregorini et al. 1994), i.e. two galaxies of roughly equal brightness
inside a common halo. Moreover, the major axes of these systems are
both roughly aligned in the East--West direction, along the main axis
of the cluster pair. 
In Fig. \ref{fig:clhisto}b and \ref{fig:clhisto}c, the black dots show
the radial velocity of each components of these dumb-bell systems.
It is interesting to note that these components are in correspondence
of two separated peaks in the velocity distributions; however, it is 
not possible to assess at a high confidence level the significance
of this bimodality (see below). 

%-------------------------------------------------------------------------------
% TABLE 3.
\setcounter{table}{2} 
\begin{table}
\caption[]{ Dynamical parameters of clusters in the A3528 complex }
\begin{flushleft}
\begin{tabular}{rrlll}
\hline\noalign{\smallskip}
Cluster & N$_{gal}$ & $<v>$ (km/s) & $\sigma$ (km/s) & notes \\
\noalign{\smallskip}
\hline\noalign{\smallskip}
 A3528 & 135 & $16332^{+72}_{-116}$ & $955^{+86}_{-86}$  & 1 Abell radius \\
 A3530 & ~54 & $16274^{+78}_{-110}$ & $730^{+143}_{-53}$  & 13' radius \\
 A3532 & ~45 & $16605^{+76}_{-161}$ & $618^{+158}_{-28}$ & 13' radius  \\ 
 A3535 & ~53 & $20192^{+78}_{-62}$ & $447^{+42}_{-22}$  & 1 Abell radius \\ 
 \noalign{\smallskip}
\hline
\end{tabular}
\end{flushleft}
\label{tab:dynamics}
\end{table}
%-------------------------------------------------------------------------------

Also in the center of A3528 there are two bright galaxies, which are located
near the positions of the two X-ray blobs observed by Schindler (1996),
already discussed in Sect. 2. In Fig. \ref{fig:clhisto}a the radial velocities
of these two galaxies are indicated by black dots: also in this case, there
is a correspondence with two peaks in the velocity distribution. 

For what concerns A3535, there is not a clear dominant galaxy near the center,
although Gregorini et al. (1994) report the presence of a dumb-bell system.    

In order to assess the statistical significance of the peaks in the velocity
distribution, we applied the KMM test (Ashman, Bird \& Zepf, 1994) which
checks whether the distribution is better represented by a single or by a
multiple Gaussian. Given the limited number of objects, the bimodality
hypothesis is not significantly better than that of a single Gaussian for all
clusters, even in the case of $b_J < 18$. However, the presence of these
peaks leads to a marginal deviation from Gaussianity in A3528 (skewness test)
and in A3530 (kurtosis test). 

%-------------------------------------------------------------------------------
% FIGURE 6 
\begin{figure}
\epsfysize=8.5cm
\epsfbox{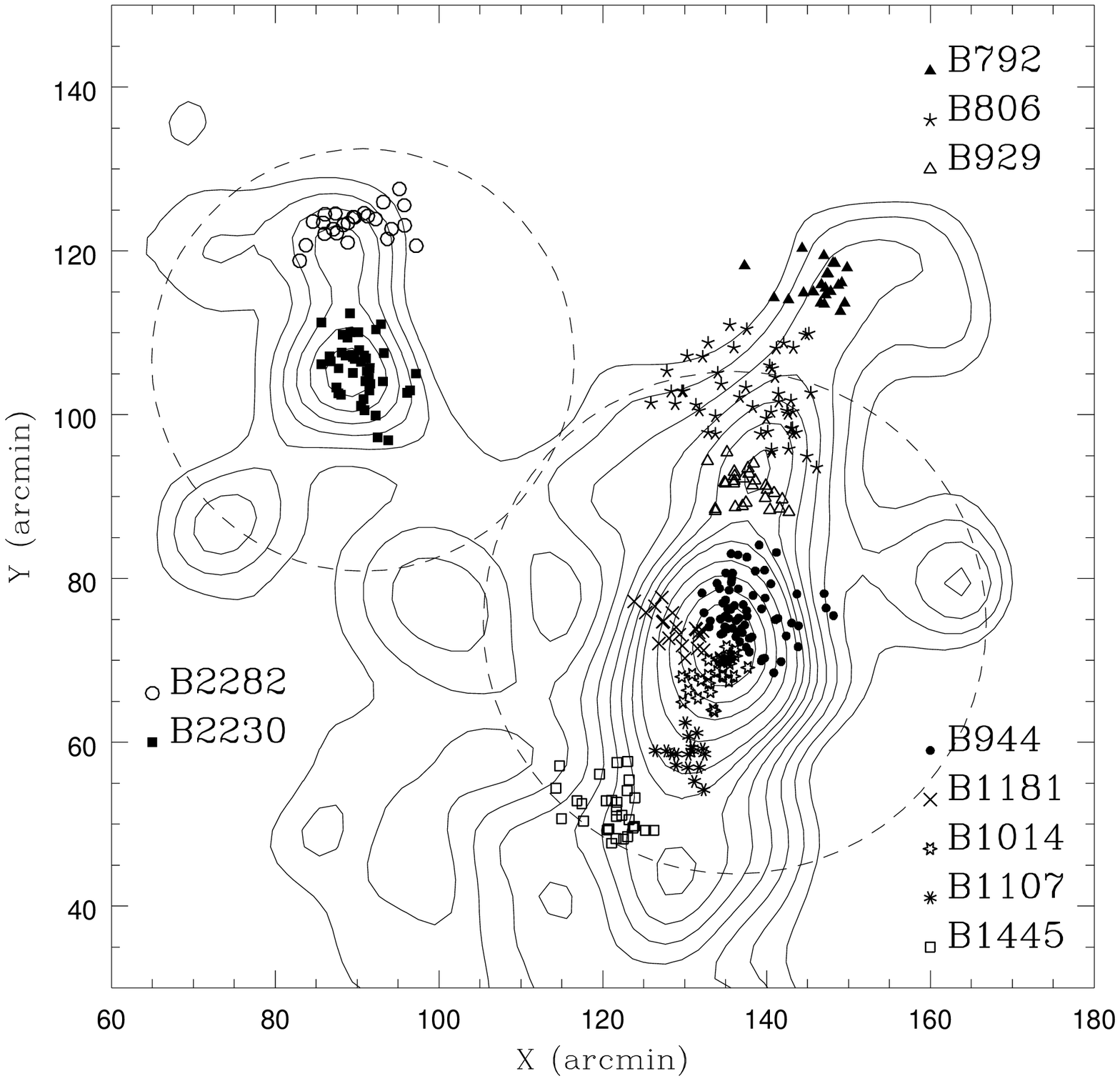}
\epsfysize=8.5cm
\epsfbox{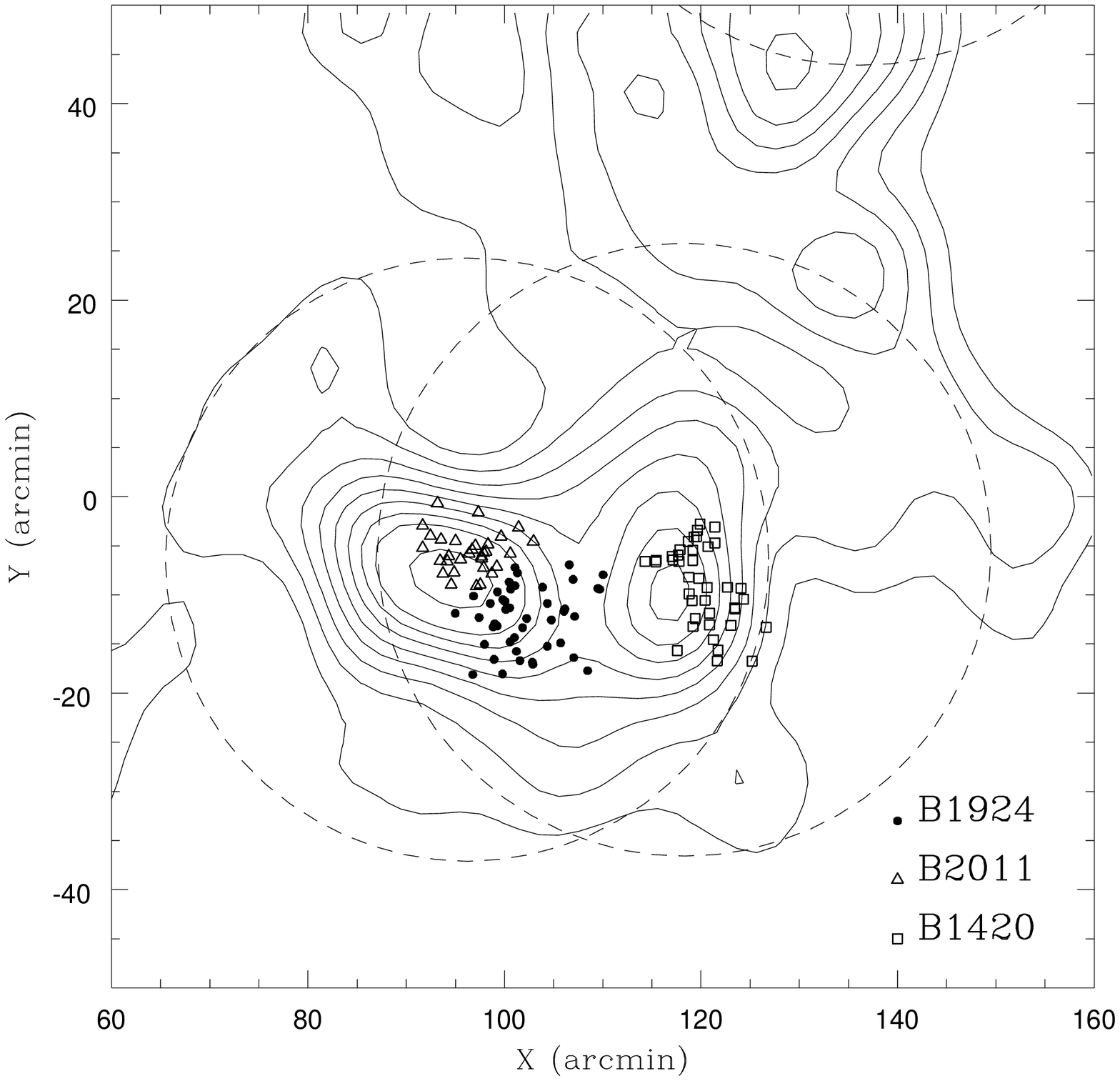}
\caption[]{ Groups in the A3528 region found in the bi--dimensional sample.
The positions of subcluster galaxies are overplotted to the 2-D isodensity
contours. Symbols used to label the different groups are explained in the 
figure. Upper panel: A3535 and A3528. Lower panel: A3532 and A3530. }
\label{fig:bi-dim}
\end{figure}
%-------------------------------------------------------------------------------

%-------------------------------------------------------------------------------
\begin{table}
\caption[]{Groups in the bi--dimensional sample }
\begin{flushleft}
\begin{tabular}{lllll}
%\hline\noalign{\smallskip}
\noalign{\smallskip}
\hline\noalign{\smallskip}
  \#  & $\alpha$ (2000) & $\delta$ (2000) & \# mem.&  notes \\
\noalign{\smallskip}
\hline\noalign{\smallskip}
 B792 & 12~53~32.6 & -28~20~00 & 22 &  ~~~~~     \\
\multicolumn{5}{l}{                                  } \\
 B806 & 12~53~53.6 & -28~34~25 & 50 &  A3528     \\
 B944 & 12~54~10.7 & -29~00~05 & 68 &  A3528     \\
 B929 & 12~54~12.1 & -28~44~56 & 27 &  A3528     \\
B1014 & 12~54~27.7 & -29~07~30 & 31 &  A3528     \\
B1107 & 12~54~44.0 & -29~17~30 & 17 &  A3528     \\
B1181 & 12~54~50.6 & -29~02~09 & 20 &  A3528     \\
B1445 & 12~55~26.3 & -29~24~09 & 29 &  A3528     \\
\multicolumn{5}{l}{                                  } \\
B1420 & 12~55~35.5 & -30~21~31 & 40 &  A3530     \\
\multicolumn{5}{l}{                                  } \\
B1924 & 12~56~58.1 & -30~28~58 & 44 &  A3532     \\
B2011 & 12~57~17.6 & -30~21~41 & 31 &  A3532     \\
\multicolumn{5}{l}{                                  } \\
B2230 & 12~57~49.2 & -28~30~27 & 43 &  A3535     \\
B2282 & 12~57~52.8 & -28~12~44 & 24 &  A3535     \\
\noalign{\smallskip}
\hline
\end{tabular}
\end{flushleft}
\label{tab:bi-dim}
\end{table}
%-------------------------------------------------------------------------------

%-------------------------------------------------------------------------------
% FIGURE 7  
\begin{figure}
\epsfysize=11.5cm
\epsfbox{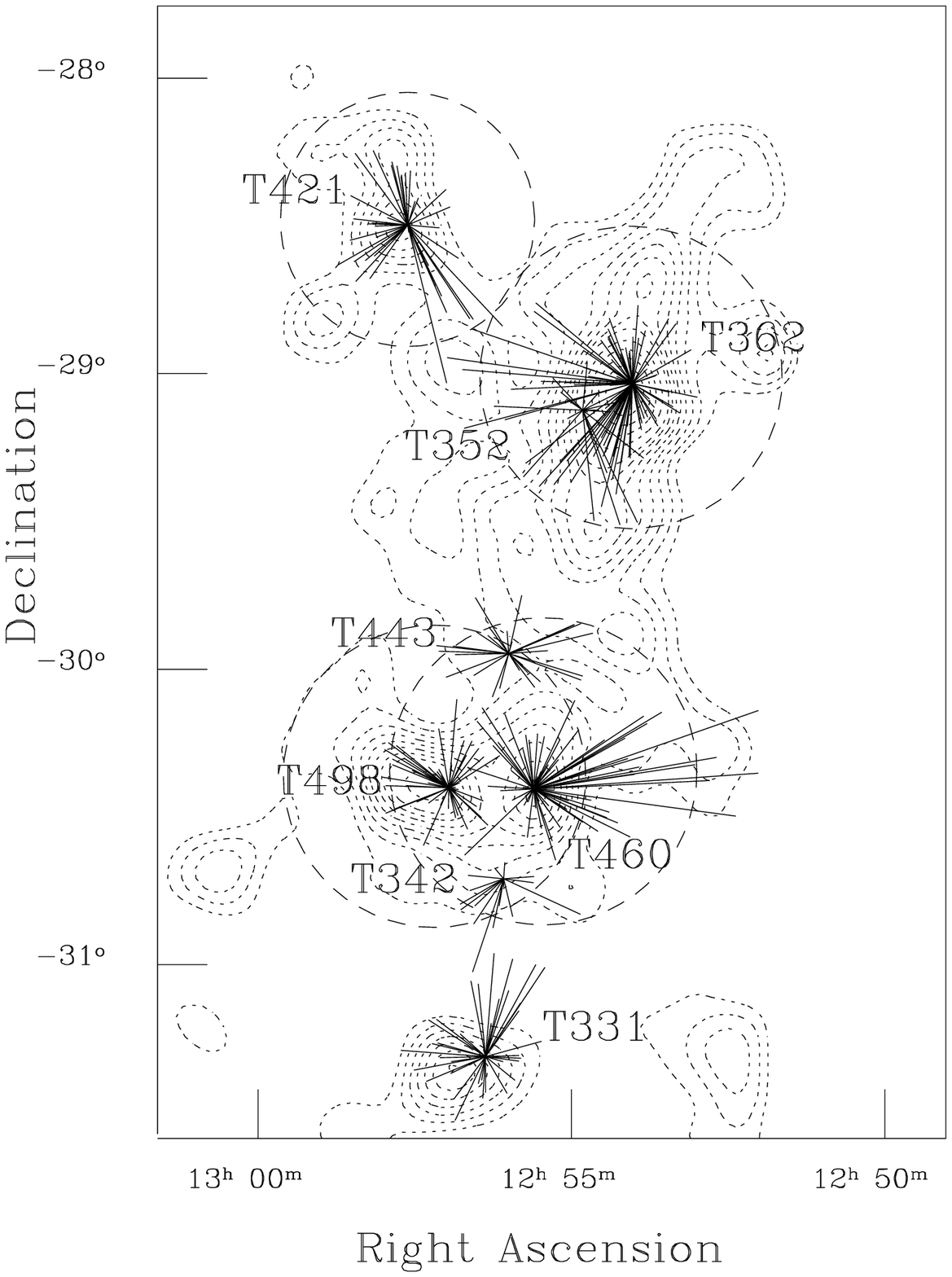}
\epsfysize=8.5cm
\epsfbox{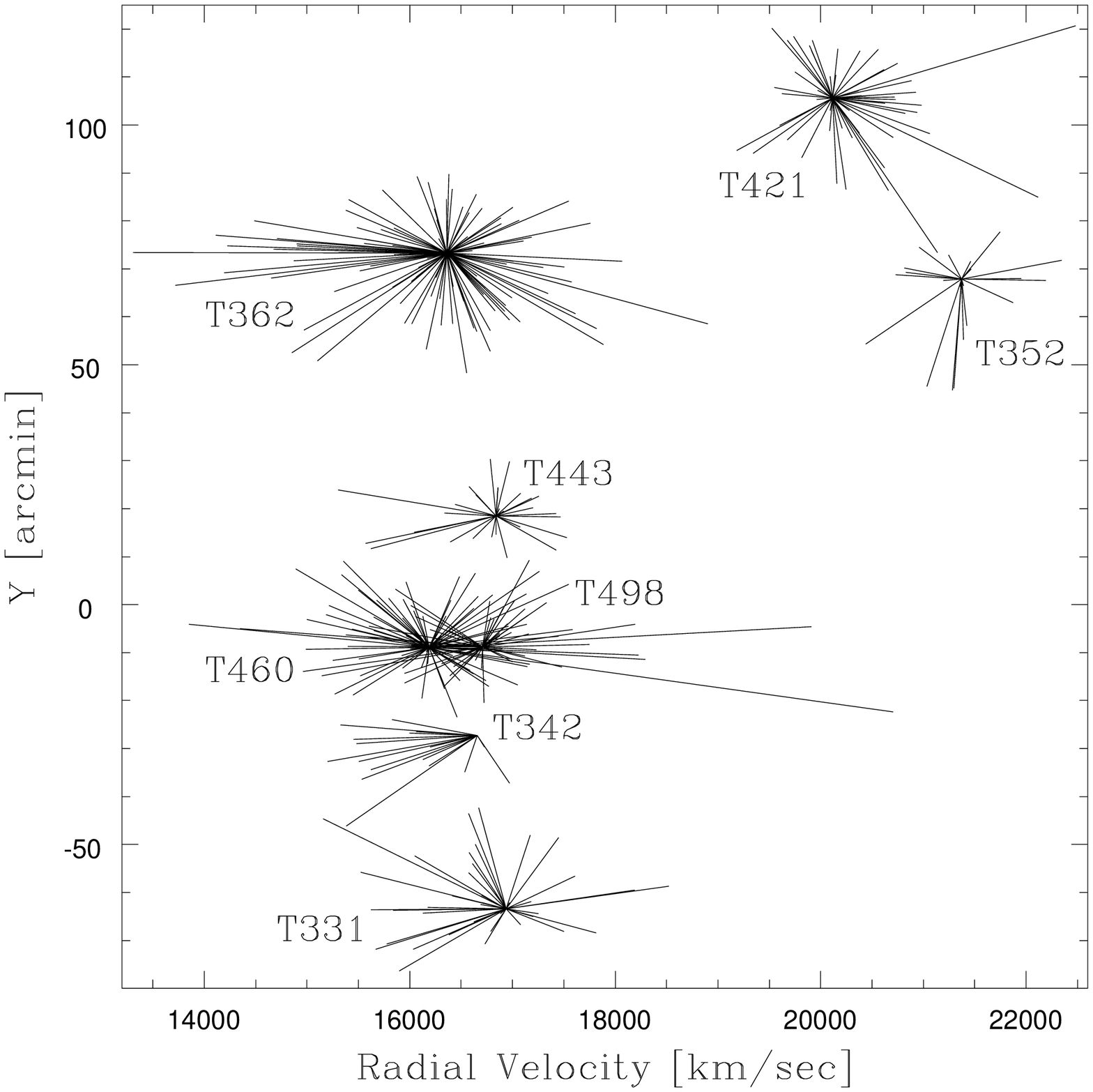}
\caption[]{ Groups in the A3528 region found in the three--dimensional sample.
Solid lines connect the position of each galaxy to the common limiting
position $\vec{x}_{lim}$ of the group. 
Upper panel: projection on the plane of the sky. Lower panel: projection on
the velocity--Y plane. }
\label{fig:three-dim}
\end{figure}
%-------------------------------------------------------------------------------

%-------------------------------------------------------------------------------
\begin{table*}
\caption[]{ Groups in the three--dimensional sample }
\begin{flushleft}
\begin{tabular}{llllllll}
%\hline\noalign{\smallskip}
\noalign{\smallskip}
\hline\noalign{\smallskip}
  \#  & $\alpha$ (2000)  & $\delta$ (2000) & $v_o$ (km/s) &\# mem. & 
  $<v>$ (km/s) & $\sigma$ (km/s) & notes \\
\noalign{\smallskip}
\hline\noalign{\smallskip}
 T362 & 12~54~36.6 & -29~05~15 & 16223 & 111 & 16313$^{+93}_{-76}$ & 
        888$^{+103}_{-82}$  & A3528          \\
 T352 & 12~54~57.3 & -29~11~47 & 21354 & ~21 & 21336$^{+115}_{-84}$ & 
        454$^{+85}_{-86}$  & A3528          \\
\multicolumn{8}{l}{                                  } \\
 T460 & 12~55~12.8 & -30~22~35 & 16322 & ~78 & 16186$^{+61}_{-97}$ & 
        637$^{+51}_{-38}$  & A3530          \\
\multicolumn{8}{l}{                                  } \\
 T443 & 12~56~08.0 & -29~57~13 & 16748 & ~27 & 16914$^{+78}_{-84}$ & 
        375$^{+77}_{-40}$  & A3530/A3532    \\
 T342 & 12~56~19.6 & -30~47~42 & 15891 & ~17 & 15852$^{+100}_{-290}$ & 
        511$^{+83}_{-84}$  & A3530/A3532    \\
\multicolumn{8}{l}{                                  } \\
 T331 & 12~56~32.6 & -31~16~24 & 16679 & ~41 & 16657$^{+84}_{-138}$ & 
        665$^{+143}_{-47}$  & ~~~~~          \\
\multicolumn{8}{l}{                                  } \\
 T498 & 12~57~07.4 & -30~23~29 & 16551 & ~48 & 16646$^{+56}_{-162}$ & 
        612$^{+93}_{-82}$  & A3532          \\
\multicolumn{8}{l}{                                  } \\
 T421 & 12~57~51.5 & -28~31~53 & 20315 & ~53 & 20239$^{+71}_{-78}$ & 
        455$^{+35}_{-45}$  & A3535          \\
\noalign{\smallskip}
\hline
\end{tabular}
\end{flushleft}
\label{tab:three-dim}
\end{table*}
%-------------------------------------------------------------------------------

%-------------------------------------------------------------------------------
%
\section{Substructure analysis}

The presence of substructures and their relevance in term of mass with respect
to the main clusters are useful tools for understanding if a structure is 
relaxed. In order to detect subcomponents, we have used the DEDICA method 
(Algorithm driven by the density estimate for the identification of clusters,
Pisani 1993, 1996), already applied by Bardelli et al. (1998b) to the
A3558 cluster complex. 
The basic idea of this algorithm is to determine the density field 
$f(\vec{x})$ (where $\vec{x}$ is the $n$-dimensional position vector)
by smoothing the distribution of galaxy with a Gaussian adaptive kernel: the  
$f(\vec{x})$ is then represented as a sum of Gaussians, whose widths
are dependent on the local density. Groups are found as local maxima 
of $f(\vec{x})$ , while a likelihood function is constructed to estimate
the confidence of the existence of the substructures (see for details
Pisani 1993, 1996).  
As done in Bardelli et al. (1998b), we performed the analysis both in the
bi--dimensional and three--dimensional samples: the first case, having more 
objects, does not suffer by small numbers statistic but could be biased by 
projection effects; the second case, although using a much smaller sample, 
is able to separate in velocity the clumps along the line of sight.

%-------------------------------------------------------------------------------
%
\subsection{The bi--dimensional sample}

The bi--dimensional sample is obtained by using the rectangular region 
described in Sect. 2, which contains 3048 galaxies brighter than $b_J=19.5$.
As done in Bardelli et al. (1998b), we estimated the background
density by plotting the histogram of the widths $\sigma$ of the Gaussians, 
which is a symmetric distribution with a tail at high $\sigma$.
We define the value of the width of the background ($\sigma_{bck}$) as 
the value at the beginning of the tail,
resulting in $\sigma_{bck}=3.85$, corresponding to a density of $0.024$
gal arcmin$^{-2}$. The whole range of densities in our sample is $0.015-1.587$
gal arcmin$^{-2}$.
The adopted $\chi^2$ threshold is 14.0, corresponding to a significance of 
$99.99\%$.

The 13 significant groups found in our bi--dimensional sample are reported in 
Table \ref{tab:bi-dim}, where 
column (1) is the identification number and columns (2) and (3) give the 
$\alpha$ and $\delta$ coordinates of the group center. These coordinates
are identified as the common $\vec{x}_{lim}$ positions of the members and 
they do not necessarily coincide with the geometrical center. Column (4) 
reports the number of the substructure members and column (5) indicates the 
main component to which the substructure belongs. Note that this association
has been made on the basis of a simple coincidence on the plane of the sky.

In Fig. \ref{fig:bi-dim} the positions of the substructure members are shown,
superimposed on the isodensity contours of galaxies. Note that in these figures 
the isodensity contours are obtained directly by binning the data (see details
in Sect. 3) and not from the $\hat{f}_k$: however the two methods give similar
contours.
Note that the objects displayed in Fig. \ref{fig:bi-dim} are the galaxies 
which resulted significantly assigned to the subclumps by our algorithm, but 
it is possible that the effective number of members of each group is higher.

The bi--dimensional analysis revealed that the A3528 complex is significantly
substructured, with the presence of two or more clumps in the inner part of  
the single clusters (with the exception of A3530).
In particular, it is interesting to note the series of clumps around A3528,
aligned along the direction pointing toward A3530-A3532.
The two X-ray blobs A3528N and A3528S are identified with B944 and B1107,
respectively, at a separation of $\sim 2.5$ arcmin and $\sim 3.9$ arcmin.  
The two dominant galaxies of A3528 (see Sect. 4) are located one in
B944 and the other in B1107. 

%-------------------------------------------------------------------------------
%
\subsection{The three--dimensional sample}

In this case we used the velocity sample described in Sect. 2:   
from the 671 galaxies with measured redshift we selected the 498 objects in
the velocity range $[10000-22500]$ km/s. 

Our formalism
uses symmetric three--dimensional Gaussian kernels, with $\sigma_x=\sigma_y=
\sigma_v$, and therefore it is necessary to scale the velocity interval
in order to have a numerical range comparable to the other two variables:
we have chosen to compress the velocity by a factor 100 (see Bardelli et al.
1998b for details about this point).

The adopted value of the background is $\sigma_{bck}=10.0$ arcmin, 
corresponding to $3.6\times 10^{-4}$ gal arcmin$^{-2}$ (km/s/100)$^{-1}$ and 
a significance threshold of $\chi^2=15.8$.
The significant 3-D clusters are 8, out of a total of 19, and are reported in 
Table \ref{tab:three-dim}. 
Column (1) is the subcluster identification number, columns (2), (3) and (4) 
report the coordinates ($\alpha$, $\delta$ and $v_o$) of the group center, 
while the number of members is reported in column (5).
The estimated dynamical parameters are reported in columns (6) and (7) and
in column (8) the association to the clusters of the structure is given.
Also in this case, the association has been made on the basis of a simple 
coincidence on the plane of the sky.
Note that $v_o$ is the projection of $\vec{x}_{lim}$ on the velocity axis, and 
therefore it does not necessarily coincide with $<v>$.  

In Fig. \ref{fig:three-dim} the members of these groups are plotted: solid
lines connect the position of each galaxy to the common limiting position 
$\vec{x}_{lim}$ of the group. In the upper panel, the groups are projected
on the plane of the sky, while in the lower panel they are shown in the
velocity--Y plane in order to clarify the separation in velocity.

In this case, the sparse sampling introduced by the redshift survey 
with respect to the bi--dimensional sample leads to the loss of some 
substructures. In particular, only the A3528 cluster remains bimodal,  
with the main component at 16313 km/s (T362) and a secondary clump at 
21336 km/s (T352). 

%-------------------------------------------------------------------------------
%
\section{Discussion}

The A3528 cluster complex is located in a region of enhanced dynamical
activity, where clusters are accreting matter in a series of spectacular 
mergings, as seen in the nearby system dominated by A3558 (see Bardelli 
et al. 1998b). However, these two cluster complexes show some relevant
differences. 

In the A3558 complex, the members of the chain appeared to be fragmented in 
a large number of components, both in optical and in X-ray band. 
In fact, from the optical substructure analysis we found 21 significant 
three-dimensional subclumps; in the X-ray band, in addition to the three ACO 
clusters A3558, A3562 and A3556, also two poor groups (SC1329-313 and 
SC1327-312) were detected. All these components are embedded in a hot gas 
filament (Bardelli et al. 1996; Kull \& B\"ohringer 1999). 
Moreover, Venturi et al. (2000) found a deficiency of radiogalaxies in the A3558
complex with respect to the radio--optical luminosity function of normal 
clusters and Bardelli et al. (1998b) found an excess of blue galaxies in the 
expected position of the shock. 
\\
These facts could suggest that the dynamical processes acting on this complex 
are in a rather advanced stage and that the merging events were already able to 
induce modifications in the galaxy population. 
Our conclusion is that the A3558 complex is a cluster-cluster collision 
(otherwise called ``major merging") seen just after the first core-core
encounter, where an intervening cluster impacted onto the richer object
A3558. Indeed the clumpiness found eastward of A3558 could be due to the
galaxies of this intervening cluster, which is now emerging from the main
component. This scenario is also confirmed by numerical simulations of
merging clusters (see f.i. lower panels in Fig.2 of Burns et al. 1994,
which show the structure after the core-core encounter).  

The A3528 complex is characterized in the X-ray band (ROSAT pointed 
observations) by two pairs of roughly similar interacting clumps: 
one pair has been resolved as two optical clusters (A3530 and A3532), 
while the two components of the other pair are both associated to A3528.
The radio--optical luminosity function of the radiogalaxies in this complex
does not appear significantly different from that of normal clusters
(Venturi et al., in preparation).   
\\
From the analysis of optical data, we find that the velocity histograms of 
cluster galaxies are not significantly different from a Gaussian distribution, 
meaning that the inner parts of the clusters are not far from the virial 
equilibrium. 
However, from our substructure analysis the A3528 complex appears characterized
by a number of subclumps. 
In particular, it is interesting to note in the bi--dimensional sample 
the series of clumps around A3528, aligned along the direction pointing toward 
A3530-A3532. Moreover, two of them appear associated with A3528N and A3528S
(B944 and B1107, respectively). The association is based not only on the 
$\alpha - \delta$ positional coincidence, but also on the mean velocity 
concordance: in fact, for each group the mode of the velocity distribution of 
members with redshift is in agreement with the velocity of the dominant galaxy
of the corresponding X-ray blob.
Also in the three--dimensional sample A3528 appears formed by two groups,
T362 and T352, but while the first is located at $<v>=16313$ km/s, the second
has $<v>=21336$ km/s, i.e. it is placed in the background of the cluster. 
In fact, it results that most members of B944 and B1107 are part of the same
three--dimensional clump (T352). 
This link of some bi--dimensional groups into a single three--dimensional
clump happens also for other substructures and is due to the sparse sampling
introduced by the redshift survey, which in some cases does not allow a 
statistically significant separation of the subcomponents.
\\
For what concerns the A3530-A3532 pair, the most puzzling feature is
the presence of two dumb-bell galaxies at their centers, elongated 
along the axis joining the two clusters (see Sect. 4). 
Moreover, it is interesting to note the presence of two subclumps (T443 and
T342) at the intersection of the Abell radii of the clusters, with mean
velocity in agreement with the main components.
\\
However, the A3528 complex is less fragmented than the A3558 structure:
in fact, even if these two complexes have similar sizes, in the A3528 complex
we detect only 8 significant three-dimensional subclumps against the 21
detected in the A3558 complex (see above). 

All these characteristics of the A3528 complex seem to indicate that this
is a ``young" structure: the two interacting cluster pairs (A3528N-A3528S and 
A3530-A3532) resemble the simulations of Burns et al. (1994) for the 
pre-merger case (see upper panels of their Fig.2). This scenario is
consistent with the suggestions of Reid et al. (1998), based on the analysis
of extended radiosources in this region and with the fact that the merging
effects on the galaxy population and on the cluster dynamics are not yet 
evident. 
\\
Given the overall overdensity of this region, at the end these two main 
components will merge together in a major merging event, forming a structure
similar to the A3558 complex.  
The ``age" difference of the two complexes is mainly due to their 
overdensities, which lead to different collapse times (see Table 5 in
Bardelli et al. 2000). This fact confirms that the Shapley Concentration
is a ``laboratory" where it is possible to study the formation of clusters
at different stages. 

A better understanding of the status of the merging process in the A3528
complex will be reached
using XMM observations of all clusters in the complex (AO1 approved 
observations), which will allow us to derive the distribution of the diffuse
hot gas and its temperature map, leading to estimate the cluster masses and 
to individuate the position of the shock fronts. 

Finally, a more detailed analysis of the galaxy population properties will
be possible through the spectral classification of galaxies in the two cluster
complexes and in the underlying supercluster ``field" (Baldi et al., in
preparation). This study will help to understand the effects of the cluster
large--scale dynamics on the galaxy morphology. 
 
%-------------------------------------------------------------------------------
%
\section*{Acknowledgements}
We warmly thank T.C.Beers, C.M.Bird and A.Pisani for having distributed
public versions of their programmes (ROSTAT, KMM and DEDICA). We also thank 
the referee (M.Drinkwater) for useful comments which improved the presentation
of the results. 

%
%-------------------------------------------------------------------------------
%

%-------------------------------------------------------------------------------
%

%
%-------------------------------------------------------------------------------
\setcounter{table}{1} 
\begin{table}
\caption[]{ Redshift data for the A3528 complex sample}
\begin{flushleft}
\begin{tabular}{rrrrrr}
\hline\noalign{\smallskip}
$\alpha$ (2000) & $\delta$ (2000) & $b_J$ & $v$ (km/s) & err & notes \\
\noalign{\smallskip}
\hline\noalign{\smallskip}
 12~51~18.66 & -30~19~02.7 & 14.98 &  4564 &  28 &       \\
 12~51~19.41 & -30~25~42.7 & 18.80 & 31465 &  63 &       \\
 12~51~23.97 & -30~18~04.1 & 15.87 &  4579 &  61 & emiss \\
 12~51~27.13 & -30~10~56.0 & 17.76 & 19390 &  15 & emiss \\
 12~51~27.89 & -30~19~00.1 & 18.56 & 27564 &  51 &       \\
 12~51~39.32 & -30~17~10.8 & 17.94 & 27326 &  46 &       \\
 12~51~40.76 & -30~27~05.8 & 18.80 & 31584 &  79 & emiss \\
 12~51~46.79 & -28~23~20.5 & 17.29 & 15854 &  57 &       \\
 12~51~50.56 & -28~23~35.7 & 17.29 & 25999 &  42 & emiss \\
 12~51~52.01 & -28~18~55.3 & 18.69 & 25735 &  95 & emiss \\
 12~52~01.51 & -30~10~46.7 & 18.73 & 20274 & 116 & emiss \\
 12~52~02.43 & -28~23~40.9 & 17.48 & 15705 &  30 &       \\
 12~52~09.63 & -28~32~25.7 & 17.22 & 16995 &  32 &       \\
 12~52~09.63 & -28~13~30.5 & 17.29 & 21211 &  81 &       \\
 12~52~12.90 & -28~20~18.7 & 18.20 & 29812 &  37 &       \\
 12~52~13.37 & -30~19~49.6 & 17.62 & 16042 & 104 &       \\
 12~52~15.12 & -30~07~06.9 & 16.87 & 17264 &  36 &       \\
 12~52~19.74 & -30~30~20.0 & 17.90 & 31572 &  68 & emiss \\
 12~52~31.14 & -28~23~24.1 & 17.71 & 14806 &  36 &       \\
 12~52~32.40 & -28~12~27.0 & 18.77 & 54096 &  15 & emiss \\
 12~52~32.98 & -28~27~14.5 & 17.90 & 17032 &  56 &       \\
 12~52~33.71 & -30~28~38.7 & 18.89 & 16618 &  15 & emiss \\
 12~52~33.92 & -30~26~52.1 & 18.97 & 60325 &  69 &       \\
 12~52~34.43 & -30~14~51.3 & 17.58 & 27457 &  15 & emiss \\
 12~52~35.82 & -30~08~16.8 & 18.86 & 54163 &  92 &       \\
 12~52~36.51 & -28~25~25.9 & 18.69 & 29047 &  64 & emiss \\
 12~52~40.05 & -28~23~38.2 & 18.78 & 14844 &  38 &       \\
 12~52~43.40 & -30~24~01.3 & 18.47 & 32472 &  67 &       \\
 12~52~45.21 & -30~35~38.9 & 18.99 & 21475 &  15 & emiss \\
 12~52~46.33 & -28~36~13.6 & 17.73 & 14886 &  45 &       \\
 12~52~46.37 & -30~13~36.5 & 17.08 & 20949 &  34 &       \\
 12~52~46.40 & -30~28~06.3 & 18.18 & 32571 &  68 &       \\
 12~52~46.95 & -30~32~57.6 & 18.66 & 31832 &  49 &       \\
 12~52~47.63 & -30~16~40.7 & 18.03 & 31450 &  48 &       \\
 12~52~48.32 & -30~06~51.4 & 18.49 & 27398 &  15 & emiss \\
 12~52~49.02 & -28~15~29.9 & 18.70 & 50037 &  60 & emiss \\
 12~52~49.37 & -30~30~18.1 & 18.32 & 31372 &  54 &       \\
 12~52~51.30 & -30~14~59.3 & 18.80 & 31876 &  95 & emiss \\
 12~52~51.59 & -30~26~59.9 & 18.76 & 31911 &  58 &       \\
 12~52~53.09 & -30~16~52.5 & 19.01 & 15650 &  15 & emiss \\
 12~52~56.46 & -29~48~28.1 & 16.73 & 15057 &  45 &       \\
 12~52~58.99 & -30~21~35.2 & 17.73 & 15739 &  57 &       \\
 12~53~00.05 & -30~09~13.3 & 18.89 & 25956 &  44 &       \\
 12~53~00.90 & -28~36~51.6 & 18.11 & 16716 & 107 &       \\
 12~53~04.40 & -30~23~54.3 & 18.62 & 50143 &  81 & emiss \\
 12~53~05.03 & -28~07~42.0 & 17.46 &  3042 &  15 & emiss \\
 12~53~06.53 & -30~16~33.2 & 18.77 & 31783 &  51 &       \\
 12~53~08.28 & -30~01~14.8 & 18.34 & 27931 &  94 & emiss \\
 12~53~12.70 & -28~28~55.6 & 18.24 & 16469 &  30 &       \\
 12~53~13.28 & -30~18~29.8 & 17.62 & 17138 &  43 &       \\
 12~53~13.58 & -28~25~47.4 & 18.38 & 17269 &  81 &       \\
 12~53~15.28 & -28~32~41.2 & 18.68 & 15795 &  96 & emiss \\
 12~53~16.20 & -29~08~46.6 & 18.71 & 41256 &  81 &       \\
 12~53~17.46 & -29~46~09.4 & 17.48 & 15086 &  56 &       \\
 12~53~17.72 & -30~24~38.2 & 18.94 & 32039 &  58 &       \\
 12~53~17.79 & -28~30~58.5 & 18.18 & 25612 & 119 & emiss \\
 12~53~19.12 & -30~30~19.3 & 17.97 & 30917 &  36 &       \\
 12~53~19.19 & -29~03~49.0 & 18.10 & 16515 &  79 &       \\
 12~53~19.91 & -30~26~58.2 & 17.75 & 30657 &  63 & emiss \\
 12~53~20.25 & -28~25~50.7 & 18.19 & 18555 &  15 & emiss \\
 12~53~21.23 & -30~18~55.4 & 18.32 & 44359 &  68 &       \\
 12~53~22.87 & -28~16~24.5 & 18.38 & 17689 &  15 & emiss \\
 12~53~23.69 & -28~20~45.5 & 17.00 & 16415 &  59 & emiss \\
\noalign{\smallskip}
\hline
\end{tabular}
\end{flushleft}
\label{tab:sample}
\end{table}
%-------------------------------------------------------------------------------

%-------------------------------------------------------------------------------
\setcounter{table}{1} 
\begin{table}
\caption[]{ cont. }
\begin{flushleft}
\begin{tabular}{rrrrrr}
\hline\noalign{\smallskip}
$\alpha$ (2000) & $\delta$ (2000) & $b_J$ & $v$ (km/s) & err & notes \\
\noalign{\smallskip}
\hline\noalign{\smallskip}
 12~53~24.18 & -28~28~47.0 & 16.02 &  8086 &  28 &       \\
 12~53~25.54 & -28~18~18.3 & 18.20 & 16287 &  15 & emiss \\
 12~53~26.02 & -28~54~11.4 & 17.84 & 16265 &  69 &       \\
 12~53~26.81 & -30~07~04.5 & 18.40 & 25871 &  62 &       \\
 12~53~27.40 & -28~18~36.6 & 17.78 & 33257 & 105 & emiss \\
 12~53~29.81 & -28~09~21.7 & 18.38 & 28874 &  15 & emiss \\
 12~53~30.69 & -28~15~52.6 & 17.15 & 16705 &  29 &       \\
 12~53~30.82 & -28~11~29.2 & 18.21 & 16988 &  15 & emiss \\
 12~53~34.04 & -30~03~21.4 & 18.79 & 41747 &  15 & emiss \\
 12~53~34.42 & -28~19~48.2 & 16.31 & 16281 &  40 &       \\
 12~53~34.59 & -28~18~58.2 & 18.60 & 16717 &  38 &       \\
 12~53~35.25 & -29~07~04.9 & 18.52 & 16016 &  50 &       \\
 12~53~35.77 & -28~15~01.0 & 17.15 & 16330 &  15 & emiss \\
 12~53~38.28 & -29~48~11.2 & 18.25 & 21069 &  15 & emiss \\
 12~53~39.82 & -30~04~36.6 & 18.72 & 21413 &  15 & emiss \\
 12~53~40.06 & -29~38~51.9 & 18.14 & 25854 &  82 & emiss \\
 12~53~40.49 & -28~48~57.9 & 17.06 & 16649 &  73 &       \\
 12~53~41.24 & -28~19~27.4 & 18.37 & 21029 &  70 & emiss \\
 12~53~41.26 & -29~05~52.8 & 18.49 & 16566 &  59 &       \\
 12~53~41.37 & -29~50~34.3 & 18.21 & 20962 &  83 & emiss \\
 12~53~45.67 & -30~07~58.7 & 17.53 & 16640 &  58 &       \\
 12~53~47.52 & -28~48~02.4 & 18.82 & 15736 &  49 & emiss \\
 12~53~47.95 & -28~14~11.9 & 17.96 & 15586 &  38 &       \\
 12~53~48.17 & -29~45~43.3 & 18.89 & 31033 &  15 & emiss \\
 12~53~48.20 & -30~00~00.7 & 18.40 & 20559 &  35 &       \\
 12~53~48.43 & -30~22~25.6 & 19.03 & 31913 &  31 &       \\
 12~53~50.29 & -30~00~23.6 & 17.72 & 20589 &  41 &       \\
 12~53~50.85 & -29~46~17.0 & 18.86 & 30917 &  15 & emiss \\
 12~53~53.51 & -29~47~14.1 & 18.61 & 30731 &  82 & emiss \\
 12~53~54.97 & -28~20~28.8 & 17.44 & 15679 &  29 &       \\
 12~53~55.22 & -29~09~51.3 & 18.80 & 16716 &  69 &       \\
 12~53~55.27 & -30~06~48.5 & 18.38 & 26007 &  34 &       \\
 12~53~57.03 & -28~59~25.8 & 18.53 & 16493 & 112 & emiss \\
 12~53~57.06 & -28~25~46.7 & 18.65 & 18030 & 171 &       \\
 12~53~57.10 & -29~08~29.9 & 18.42 & 17606 &  36 & emiss \\
 12~53~57.49 & -30~22~39.3 & 18.48 & 21633 &  43 &       \\
 12~53~57.89 & -29~51~46.4 & 17.92 & 27888 &  31 &       \\
 12~53~58.54 & -28~59~37.5 & 18.00 & 15833 &  38 &       \\
 12~53~58.65 & -29~06~05.5 & 18.84 & 16654 &  56 &       \\
 12~53~59.03 & -30~10~21.0 & 19.12 & 17546 &  61 &       \\
 12~54~00.50 & -29~53~58.9 & 18.73 & 27838 &  88 &       \\
 12~54~01.64 & -29~49~06.8 & 18.57 & 27762 &  73 &       \\
 12~54~01.75 & -30~16~58.5 & 18.95 & 16354 &  80 &       \\
 12~54~02.55 & -30~08~48.0 & 18.59 & 16483 &  31 &       \\
 12~54~02.66 & -30~18~46.5 & 19.19 & 20786 &  37 &       \\
 12~54~03.82 & -29~31~13.6 & 18.21 & 17018 &  38 &       \\
 12~54~04.68 & -29~08~04.0 & 18.87 & 13720 &  85 &       \\
 12~54~05.77 & -30~19~08.4 & 18.12 & 16962 &  45 &       \\
 12~54~06.21 & -28~58~19.0 & 19.34 & 16283 &  92 &       \\
 12~54~06.55 & -30~08~46.3 & 18.61 & 32535 & 105 &       \\
 12~54~06.93 & -29~50~44.6 & 18.13 & 21029 &  15 & emiss \\
 12~54~07.59 & -29~16~27.8 & 18.02 & 21424 &  71 &       \\
 12~54~07.97 & -29~08~23.1 & 18.51 & 16421 &  58 &       \\
 12~54~08.46 & -28~50~31.5 & 17.93 & 17546 &  62 &       \\
 12~54~08.69 & -29~47~51.9 & 17.16 & 21025 &  32 &       \\
 12~54~09.79 & -30~20~42.2 & 17.49 & 16046 &  53 &       \\
 12~54~09.86 & -30~16~09.0 & 19.03 & 46420 &  75 & emiss \\
 12~54~12.02 & -29~29~31.4 & 19.24 & 21296 &  10 & emiss \\
 12~54~12.05 & -29~09~22.7 & 17.64 & 15910 &  36 &       \\
 12~54~12.36 & -29~32~30.7 & 18.78 & 44170 &  93 &       \\
 12~54~12.60 & -29~03~37.5 & 18.57 & 16945 &  38 &       \\
 12~54~12.76 & -29~58~29.7 & 18.94 & 31862 &  42 &       \\
 12~54~12.80 & -29~53~04.6 & 18.04 & 21495 &  31 &       \\
\noalign{\smallskip}
\hline
\end{tabular}
\end{flushleft}
\end{table}
%-------------------------------------------------------------------------------

%-------------------------------------------------------------------------------
\setcounter{table}{1} 
\begin{table}
\caption[]{ cont. }
\begin{flushleft}
\begin{tabular}{rrrrrr}
\hline\noalign{\smallskip}
$\alpha$ (2000) & $\delta$ (2000) & $b_J$ & $v$ (km/s) & err & notes \\
\noalign{\smallskip}
\hline\noalign{\smallskip}
 12~54~12.80 & -30~33~21.8 & 16.82 & 15271 &  35 &       \\
 12~54~13.27 & -29~05~30.7 & 18.86 & 14192 &  57 &       \\
 12~54~14.35 & -29~03~02.5 & 18.10 & 18069 & 111 &       \\
 12~54~15.08 & -28~52~02.9 & 18.66 & 16706 &  57 &       \\
 12~54~15.13 & -29~49~00.1 & 17.36 & 20862 &  27 &       \\
 12~54~16.36 & -29~47~35.8 & 18.57 & 27605 &  72 & emiss \\
 12~54~16.39 & -30~17~28.5 & 18.06 & 15929 &  72 &       \\
 12~54~16.78 & -29~01~17.4 & 18.23 & 15310 &  39 &       \\
 12~54~17.36 & -29~31~03.1 & 16.96 & 24727 &  34 &       \\
 12~54~17.46 & -29~00~48.4 & 18.45 & 16359 &  51 &       \\
 12~54~18.94 & -28~59~30.3 & 18.36 & 14898 &  87 &       \\
 12~54~19.17 & -30~14~41.1 & 19.01 & 31726 &  91 & emiss \\
 12~54~19.55 & -29~01~20.0 & 18.73 & 16112 &  47 &       \\
 12~54~19.72 & -30~13~25.3 & 18.01 & 15795 &  41 &       \\
 12~54~19.99 & -29~27~40.8 & 18.84 & 24422 &  42 &       \\
 12~54~20.09 & -28~51~47.0 & 18.39 & 16515 &  41 &       \\
 12~54~20.28 & -29~04~08.1 & 16.88 & 16521 &  74 &       \\
 12~54~20.36 & -29~01~47.1 & 18.09 & 21244 &  98 &       \\
 12~54~20.38 & -29~16~27.2 & 18.42 & 16307 &  60 &       \\
 12~54~20.40 & -29~48~40.0 & 17.92 & 20828 &  60 &       \\
 12~54~20.94 & -29~06~37.6 & 18.60 & 14653 &  52 &       \\
 12~54~21.21 & -29~16~02.4 & 18.55 & 15951 &  59 &       \\
 12~54~21.52 & -28~58~00.0 & 18.76 & 16564 &  52 &       \\
 12~54~23.13 & -29~46~19.9 & 18.96 & 31914 &  77 &       \\
 12~54~23.43 & -29~05~04.0 & 17.03 & 15514 &  59 &       \\
 12~54~23.48 & -29~10~02.2 & 17.75 & 16518 &  38 &       \\
 12~54~23.50 & -28~55~10.6 & 18.76 & 17758 &  50 &       \\
 12~54~23.70 & -29~20~03.8 & 19.10 & 24521 &  92 &       \\
 12~54~24.02 & -29~34~59.0 & 19.12 & 15040 & 100 &       \\
 12~54~24.43 & -30~15~07.4 & 18.57 & 31759 &  54 &       \\
 12~54~24.62 & -28~56~07.6 & 18.73 & 15848 &  52 &       \\
 12~54~24.86 & -30~30~36.3 & 16.77 & 16742 &  37 &       \\
 12~54~24.99 & -29~04~41.1 & 18.33 & 21468 &  51 &       \\
 12~54~25.31 & -29~03~03.4 & 18.10 & 17011 &  49 &       \\
 12~54~25.59 & -29~29~21.6 & 17.89 & 28420 &  55 &       \\
 12~54~26.60 & -29~00~36.5 & 18.19 & 15869 &  56 &       \\
 12~54~26.63 & -28~59~20.9 & 18.64 & 15556 &  54 &       \\
 12~54~26.71 & -28~54~02.1 & 18.85 & 16887 &  63 &       \\
 12~54~28.77 & -29~30~00.4 & 17.98 & 21282 &  15 & emiss \\
 12~54~28.79 & -29~17~31.4 & 17.29 & 14969 &  45 &       \\
 12~54~29.54 & -29~04~26.3 & 18.92 & 16087 &  57 &       \\
 12~54~31.46 & -28~55~18.8 & 17.56 & 16892 &  45 &       \\
 12~54~31.47 & -29~10~56.8 & 18.77 & 16290 &  54 &       \\
 12~54~32.38 & -29~26~05.3 & 18.85 & 28167 &  15 & emiss \\
 12~54~32.97 & -29~23~53.0 & 19.07 & 15100 &  48 &       \\
 12~54~33.03 & -30~28~30.6 & 19.01 & 16589 &  50 &       \\
 12~54~33.16 & -28~46~29.5 & 18.68 & 55142 &  49 &       \\
 12~54~33.80 & -29~08~35.3 & 17.24 & 15835 &  49 &       \\
 12~54~34.67 & -29~59~16.6 & 18.11 & 23000 &  73 & emiss \\
 12~54~34.71 & -28~59~54.3 & 18.49 & 14224 &  42 &       \\
 12~54~34.77 & -29~23~02.9 & 19.23 & 25953 &  44 &       \\
 12~54~35.01 & -30~29~39.9 & 17.83 & 15517 &  48 & emiss \\
 12~54~35.02 & -29~07~29.2 & 18.80 & 17577 &  60 &       \\
 12~54~35.46 & -29~38~19.4 & 18.57 & 42043 &  39 &       \\
 12~54~36.39 & -29~16~12.8 & 19.17 & 16418 &  42 &       \\
 12~54~36.48 & -28~51~25.7 & 18.50 & 40883 &  56 &       \\
 12~54~36.55 & -29~20~33.5 & 18.14 & 17885 &  35 &       \\
 12~54~38.10 & -29~39~00.2 & 18.80 & 17626 &  56 &       \\
 12~54~38.31 & -30~09~19.6 & 17.94 &  4999 &  13 & emiss \\
 12~54~38.76 & -30~13~50.8 & 17.82 & 15676 &  15 & emiss \\
 12~54~38.92 & -29~01~10.6 & 18.21 & 15642 &  56 &       \\
 12~54~39.26 & -29~17~51.5 & 19.14 & 16655 &  48 &       \\
 12~54~40.25 & -29~43~52.8 & 18.60 & 16069 &  42 &       \\
\noalign{\smallskip}
\hline
\end{tabular}
\end{flushleft}
\end{table}
%-------------------------------------------------------------------------------

%-------------------------------------------------------------------------------
\setcounter{table}{1} 
\begin{table}
\caption[]{ cont. }
\begin{flushleft}
\begin{tabular}{rrrrrr}
\hline\noalign{\smallskip}
$\alpha$ (2000) & $\delta$ (2000) & $b_J$ & $v$ (km/s) & err & notes \\
\noalign{\smallskip}
\hline\noalign{\smallskip}
 12~54~40.91 & -28~53~01.6 & 17.28 & 16576 &  33 &       \\
 12~54~40.96 & -29~01~23.5 & 18.91 & 13307 &  72 &       \\
 12~54~41.84 & -29~19~33.1 & 18.33 & 21393 &  16 & emiss \\
 12~54~42.34 & -28~48~07.4 & 18.65 & 16413 &  53 &       \\
 12~54~42.41 & -28~51~55.0 & 17.48 & 17003 &  53 &       \\
 12~54~43.00 & -28~52~15.1 & 17.03 & 16242 &  42 &       \\
 12~54~43.12 & -28~50~15.9 & 17.67 & 16359 &  50 &       \\
 12~54~43.26 & -29~15~51.7 & 18.61 & 13802 &  55 &       \\
 12~54~43.39 & -29~15~25.5 & 18.62 & 16971 &  57 &       \\
 12~54~43.40 & -30~32~21.8 & 18.13 & 16340 &  33 &       \\
 12~54~43.83 & -29~30~38.8 & 18.84 & 24324 &  40 &       \\
 12~54~44.24 & -29~36~58.6 & 18.47 & 15130 & 132 &       \\
 12~54~44.26 & -29~06~30.2 & 18.34 & 16063 &  43 &       \\
 12~54~45.00 & -29~31~09.3 & 19.07 & 32275 & 106 &       \\
 12~54~45.52 & -29~14~06.2 & 18.52 & 17612 &  48 &       \\
 12~54~45.52 & -29~16~15.7 & 18.96 & 16022 &  35 &       \\
 12~54~45.65 & -29~17~54.8 & 18.21 & 13951 &  33 &       \\
 12~54~46.11 & -29~29~23.3 & 18.38 & 16269 &  27 &       \\
 12~54~46.29 & -29~26~28.8 & 18.99 & 16554 &  32 &       \\
 12~54~47.59 & -29~12~25.7 & 18.07 & 16934 &  52 &       \\
 12~54~48.04 & -30~16~56.6 & 19.25 & 15184 &  45 &       \\
 12~54~49.30 & -30~19~34.1 & 17.87 & 16081 &  34 &       \\
 12~54~49.68 & -28~52~35.5 & 18.77 & 15377 &  51 &       \\
 12~54~50.08 & -29~03~01.8 & 18.22 & 22345 & 106 &       \\
 12~54~50.39 & -30~26~08.8 & 19.10 & 16865 &  26 &       \\
 12~54~51.70 & -29~52~11.2 & 17.97 & 17259 &  34 &       \\
 12~54~52.24 & -29~17~37.8 & 18.17 & 14371 &  34 &       \\
 12~54~52.68 & -29~29~22.0 & 18.27 & 21035 & 106 &       \\
 12~54~53.60 & -29~34~43.9 & 17.30 & 16353 &  35 &       \\
 12~54~53.81 & -30~16~11.0 & 18.88 & 16039 &  68 &       \\
 12~54~53.81 & -29~00~46.6 & 17.46 & 16140 &  36 &       \\
 12~54~54.71 & -29~16~16.0 & 18.92 & 18902 &  30 &       \\
 12~54~54.83 & -30~21~34.0 & 19.02 & 15586 &  46 &       \\
 12~54~55.33 & -29~30~12.2 & 17.52 & 16324 &  37 &       \\
 12~54~55.41 & -28~54~17.0 & 17.15 & 22905 &  51 &       \\
 12~54~56.65 & -29~50~10.2 & 17.45 & 14860 &  94 & emiss \\
 12~54~57.01 & -29~11~54.3 & 18.11 & 21872 &  33 &       \\
 12~54~57.69 & -30~30~40.9 & 19.24 & 16488 &  44 &       \\
 12~54~57.85 & -29~15~55.7 & 19.10 & 14746 & 107 &       \\
 12~54~58.14 & -30~13~43.6 & 18.22 & 25943 &  81 &       \\
 12~54~58.40 & -30~49~18.6 & 18.53 & 15622 &  55 &       \\
 12~54~58.51 & -29~41~44.5 & 18.33 & 17734 & 172 &       \\
 12~55~00.18 & -29~11~33.0 & 18.64 & 16526 &  67 &       \\
 12~55~00.41 & -30~00~56.0 & 18.64 & 17528 &  35 &       \\
 12~55~00.54 & -29~07~15.1 & 18.49 & 21193 &  54 &       \\
 12~55~00.97 & -29~00~05.0 & 18.31 & 14906 & 139 &       \\
 12~55~01.04 & -30~43~37.9 & 17.69 & 42778 &  61 & emiss \\
 12~55~01.19 & -29~47~20.7 & 17.38 & 20802 &  84 & emiss \\
 12~55~02.05 & -28~57~10.6 & 17.47 & 21747 &  57 &       \\
 12~55~02.12 & -30~08~43.7 & 18.33 & 20723 &  86 &       \\
 12~55~02.26 & -30~01~47.3 & 18.72 & 16395 &  28 &       \\
 12~55~02.39 & -29~28~59.1 & 18.39 & 17692 &  45 &       \\
 12~55~02.87 & -29~50~18.8 & 18.77 & 16576 &  81 &       \\
 12~55~03.16 & -29~38~52.0 & 17.91 & 24603 &  41 &       \\
 12~55~03.30 & -30~11~50.0 & 17.52 & 15498 &  54 &       \\
 12~55~04.06 & -29~15~54.5 & 17.96 & 17076 &  37 &       \\
 12~55~04.10 & -29~25~37.5 & 17.82 & 28989 &  27 &       \\
 12~55~04.96 & -29~08~24.4 & 18.89 & 21393 &  52 & emiss \\
 12~55~05.87 & -28~58~15.9 & 18.18 & 17186 &  87 &       \\
 12~55~06.22 & -28~54~43.4 & 18.27 & 17065 & 162 &       \\
 12~55~07.29 & -30~05~52.8 & 17.49 & 15355 &  93 & emiss \\
 12~55~07.41 & -29~31~33.0 & 18.02 & 24741 &  46 &       \\
 12~55~07.87 & -29~29~44.8 & 17.92 & 17103 &  35 &       \\
\noalign{\smallskip}
\hline
\end{tabular}
\end{flushleft}
\end{table}
%-------------------------------------------------------------------------------

%-------------------------------------------------------------------------------
\setcounter{table}{1} 
\begin{table}
\caption[]{ cont. }
\begin{flushleft}
\begin{tabular}{rrrrrr}
\hline\noalign{\smallskip}
$\alpha$ (2000) & $\delta$ (2000) & $b_J$ & $v$ (km/s) & err & notes \\
\noalign{\smallskip}
\hline\noalign{\smallskip}
 12~55~08.09 & -30~45~06.4 & 18.52 & 16517 &  32 &       \\
 12~55~08.38 & -30~43~25.0 & 16.35 & 17709 &  24 &       \\
 12~55~09.61 & -29~50~31.9 & 18.08 & 16857 &  35 &       \\
 12~55~09.72 & -29~30~54.2 & 18.52 & 15163 &  50 &       \\
 12~55~11.36 & -30~26~23.2 & 18.01 & 16034 &  48 &       \\
 12~55~13.07 & -30~47~19.8 & 18.86 & 49439 &  15 & emiss \\
 12~55~13.18 & -29~40~37.1 & 17.84 & 15671 &  46 &       \\
 12~55~13.20 & -30~28~02.1 & 17.76 & 16112 &  47 &       \\
 12~55~13.73 & -29~01~48.3 & 18.72 & 15903 &  53 &       \\
 12~55~14.64 & -30~43~42.1 & 18.77 & 49665 &  15 & emiss \\
 12~55~14.96 & -29~21~42.9 & 18.24 & 16163 &  29 &       \\
 12~55~15.01 & -29~25~10.8 & 19.02 & 24705 &  36 &       \\
 12~55~16.26 & -29~33~12.8 & 19.19 & 24584 &  69 &       \\
 12~55~17.00 & -29~50~44.7 & 18.84 & 31938 &  57 &       \\
 12~55~17.99 & -29~24~22.8 & 18.74 & 41819 &  44 &       \\
 12~55~18.45 & -29~52~47.1 & 18.73 & 31229 &  34 &       \\
 12~55~18.69 & -29~26~28.1 & 16.98 & 24697 &  31 &       \\
 12~55~19.23 & -30~31~42.4 & 18.88 & 17052 &  51 &       \\
 12~55~19.37 & -29~20~51.0 & 19.21 & 24541 &  47 &       \\
 12~55~19.37 & -29~08~43.2 & 18.23 & 21359 &  90 & emiss \\
 12~55~19.47 & -29~17~17.6 & 17.76 & 16623 &  33 &       \\
 12~55~19.77 & -29~35~50.6 & 19.15 & 17876 &  43 &       \\
 12~55~20.23 & -29~51~28.6 & 18.61 & 31156 &  36 &       \\
 12~55~20.81 & -30~13~53.1 & 17.78 & 15934 &  33 &       \\
 12~55~20.95 & -29~04~44.5 & 18.54 & 20813 &  41 &       \\
 12~55~21.61 & -29~52~30.8 & 18.42 & 25973 &  36 &       \\
 12~55~21.62 & -30~18~04.1 & 18.79 & 15000 &  44 &       \\
 12~55~22.22 & -29~23~52.8 & 18.05 & 28785 &  43 &       \\
 12~55~22.82 & -29~01~24.8 & 18.19 & 16319 &  81 &       \\
 12~55~22.89 & -29~45~34.0 & 18.54 & 38722 &  73 &       \\
 12~55~22.90 & -30~38~27.9 & 18.75 & 16459 &  69 &       \\
 12~55~23.30 & -30~26~52.2 & 19.22 & 23093 &  44 &       \\
 12~55~23.88 & -30~35~06.8 & 17.16 & 23076 &  80 &       \\
 12~55~24.71 & -30~13~49.3 & 19.14 & 31800 &  32 &       \\
 12~55~24.71 & -30~20~05.3 & 18.75 & 15440 &  55 &       \\
 12~55~24.74 & -30~24~14.4 & 19.15 & 16900 &  31 &       \\
 12~55~25.02 & -29~22~17.3 & 17.67 & 24441 &  34 &       \\
 12~55~25.14 & -29~24~01.0 & 18.22 & 28949 &  50 &       \\
 12~55~25.23 & -29~26~44.7 & 18.61 & 41776 &  43 &       \\
 12~55~25.26 & -29~23~15.6 & 19.03 & 28975 &  52 &       \\
 12~55~25.35 & -29~17~26.6 & 18.39 & 17819 &  39 &       \\
 12~55~25.50 & -30~25~33.7 & 17.53 & 18229 &  54 &       \\
 12~55~26.18 & -29~33~32.0 & 18.06 & 16159 &  41 &       \\
 12~55~26.18 & -29~03~23.7 & 17.37 & 21462 &  50 &       \\
 12~55~26.31 & -30~34~31.4 & 17.69 & 16118 &  35 &       \\
 12~55~27.59 & -29~27~18.5 & 19.04 & 15425 &  31 &       \\
 12~55~30.04 & -29~25~43.2 & 18.94 & 28756 &  84 &       \\
 12~55~30.05 & -30~27~24.9 & 19.02 & 17170 &  31 &       \\
 12~55~30.47 & -29~40~29.7 & 18.87 & 20398 &  76 & emiss \\
 12~55~30.54 & -30~12~35.8 & 18.96 & 31886 &  80 &       \\
 12~55~30.87 & -31~00~16.2 & 18.96 & 16707 &  71 &       \\
 12~55~30.99 & -30~28~15.3 & 18.85 & 15933 &  75 & emiss \\
 12~55~31.01 & -29~22~09.3 & 17.47 & 16786 &  35 &       \\
 12~55~31.40 & -30~19~07.3 & 16.87 & 13849 &  41 &       \\
 12~55~31.72 & -30~21~30.5 & 18.50 & 16308 &  33 &       \\
 12~55~31.74 & -30~25~37.6 & 18.50 & 15922 &  51 &       \\
 12~55~32.99 & -31~15~18.7 & 17.25 & 16834 &  15 & emiss \\
 12~55~33.11 & -30~24~55.1 & 18.84 & 16996 &  15 & emiss \\
 12~55~33.60 & -29~55~57.1 & 17.88 & 16341 &  30 &       \\
 12~55~33.72 & -29~00~32.1 & 17.62 & 20961 &  47 &       \\
 12~55~33.98 & -29~32~48.0 & 19.21 & 24486 &  16 & emiss \\
 12~55~35.13 & -29~18~55.5 & 19.28 & 28877 &  15 & emiss \\
 12~55~35.58 & -30~13~18.9 & 17.48 & 16667 &  45 &       \\
\noalign{\smallskip}
\hline
\end{tabular}
\end{flushleft}
\end{table}
%-------------------------------------------------------------------------------

%-------------------------------------------------------------------------------
\setcounter{table}{1} 
\begin{table}
\caption[]{ cont. }
\begin{flushleft}
\begin{tabular}{rrrrrr}
\hline\noalign{\smallskip}
$\alpha$ (2000) & $\delta$ (2000) & $b_J$ & $v$ (km/s) & err & notes \\
\noalign{\smallskip}
\hline\noalign{\smallskip}
 12~55~36.44 & -30~32~03.0 & 18.01 & 16768 &  35 &       \\
 12~55~37.90 & -30~20~28.2 & 19.09 & 15649 &  60 &       \\
 12~55~39.03 & -30~26~30.9 & 18.88 & 18294 &  38 &       \\
 12~55~39.28 & -30~59~43.0 & 18.11 & 15156 &  79 &       \\
 12~55~39.89 & -29~03~32.7 & 18.84 & 30808 & 114 &       \\
 12~55~40.03 & -30~02~16.4 & 17.52 & 15568 &  38 &       \\
 12~55~40.15 & -28~49~43.0 & 17.45 & 22686 &  45 &       \\
 12~55~42.10 & -30~12~49.5 & 19.04 & 17134 &  36 &       \\
 12~55~42.41 & -30~27~58.2 & 17.93 & 17147 &  39 &       \\
 12~55~42.43 & -30~41~53.8 & 18.53 & 15997 &  33 &       \\
 12~55~42.47 & -29~33~17.2 & 17.17 &  7370 &  44 &       \\
 12~55~42.72 & -29~30~26.1 & 18.03 & 16375 &  29 &       \\
 12~55~42.95 & -30~43~06.9 & 16.16 & 15451 &  39 &       \\
 12~55~43.47 & -29~24~41.3 & 17.02 & 28927 &  73 &       \\
 12~55~44.30 & -30~24~19.0 & 17.84 & 16004 &  46 &       \\
 12~55~44.64 & -29~22~32.2 & 18.86 & 14855 &  90 & emiss \\
 12~55~44.72 & -29~01~12.0 & 18.19 & 16611 &  57 &       \\
 12~55~45.60 & -30~24~52.9 & 18.32 & 16162 &  75 &       \\
 12~55~45.62 & -30~03~24.3 & 18.19 & 15623 &  60 &       \\
 12~55~48.18 & -30~23~23.6 & 18.78 & 16652 &  50 &       \\
 12~55~48.90 & -30~21~33.6 & 17.91 & 16536 &  32 &       \\
 12~55~49.45 & -30~33~58.8 & 18.87 & 15449 &  60 &       \\
 12~55~49.59 & -30~14~08.0 & 17.44 & 16398 &  52 &       \\
 12~55~49.90 & -30~26~44.3 & 18.30 & 15250 &  51 &       \\
 12~55~50.03 & -30~39~37.0 & 18.74 & 17097 &  40 &       \\
 12~55~50.74 & -28~45~15.8 & 18.02 & 16381 &  40 &       \\
 12~55~51.71 & -31~18~51.0 & 17.58 & 15839 &  32 &       \\
 12~55~51.83 & -30~00~28.4 & 18.17 & 16840 &  28 &       \\
 12~55~51.96 & -28~46~59.6 & 18.35 & 16178 &  15 & emiss \\
 12~55~52.53 & -29~41~29.2 & 19.10 & 16132 &  41 &       \\
 12~55~52.67 & -30~25~32.6 & 18.21 & 16274 &  31 &       \\
 12~55~53.25 & -30~13~04.7 & 18.85 & 16808 &  56 &       \\
 12~55~53.32 & -31~07~31.5 & 18.60 & 16048 &  33 &       \\
 12~55~53.48 & -30~27~25.0 & 17.32 & 16174 &  29 &       \\
 12~55~53.93 & -30~17~28.4 & 19.04 & 16130 &  15 & emiss \\
 12~55~54.01 & -31~18~12.3 & 18.48 & 16176 &  40 &       \\
 12~55~54.04 & -31~19~27.2 & 17.14 & 16128 &  34 &       \\
 12~55~54.18 & -30~21~42.0 & 18.52 & 17451 &  33 &       \\
 12~55~54.50 & -31~09~09.3 & 17.38 & 16609 &  34 &       \\
 12~55~55.03 & -30~23~47.0 & 18.79 & 15825 &  44 &       \\
 12~55~55.06 & -30~43~17.2 & 16.20 & 16753 &  30 &       \\
 12~55~55.35 & -30~15~54.4 & 18.03 & 16623 &  32 &       \\
 12~55~56.07 & -29~07~15.4 & 18.68 & 31080 &  66 & emiss \\
 12~55~56.82 & -29~58~58.6 & 18.09 & 17073 &  35 &       \\
 12~55~57.02 & -29~55~40.0 & 18.41 & 16915 &  37 &       \\
 12~55~57.17 & -31~23~15.8 & 17.66 & 16790 &  44 &       \\
 12~55~57.52 & -29~17~58.4 & 17.15 & 16784 &  41 &       \\
 12~55~58.55 & -30~25~43.1 & 17.92 & 16418 &  29 &       \\
 12~55~58.95 & -29~44~48.2 & 17.87 & 16785 &  62 &       \\
 12~55~59.37 & -29~20~43.5 & 18.78 & 20440 &  75 &       \\
 12~55~59.82 & -29~39~39.0 & 19.23 & 16486 &  77 &       \\
 12~55~59.92 & -30~00~01.9 & 17.78 & 16044 &  15 & emiss \\
 12~56~00.28 & -29~35~47.3 & 18.83 & 16099 &  40 &       \\
 12~56~01.55 & -30~58~42.3 & 18.72 & 16573 &  38 &       \\
 12~56~02.14 & -29~36~10.7 & 18.95 & 16107 &  81 &       \\
 12~56~02.49 & -30~50~04.6 & 18.34 & 16535 &  51 &       \\
 12~56~03.76 & -29~42~52.7 & 17.20 &  4882 &  66 & emiss \\
 12~56~04.36 & -30~42~02.2 & 18.13 & 17022 &  37 &       \\
 12~56~04.46 & -30~26~27.8 & 18.30 & 15503 &  39 &       \\
 12~56~04.69 & -31~05~06.0 & 18.35 & 16640 &  39 &       \\
 12~56~04.70 & -31~25~57.8 & 18.49 & 16734 &  85 &       \\
 12~56~05.40 & -29~27~23.1 & 17.88 & 25628 &  33 &       \\
 12~56~06.29 & -30~20~19.9 & 18.61 & 17586 &  37 &       \\
\noalign{\smallskip}
\hline
\end{tabular}
\end{flushleft}
\end{table}
%-------------------------------------------------------------------------------

%-------------------------------------------------------------------------------
\setcounter{table}{1} 
\begin{table}
\caption[]{ cont. }
\begin{flushleft}
\begin{tabular}{rrrrrr}
\hline\noalign{\smallskip}
$\alpha$ (2000) & $\delta$ (2000) & $b_J$ & $v$ (km/s) & err & notes \\
\noalign{\smallskip}
\hline\noalign{\smallskip}
 12~56~07.07 & -29~54~15.4 & 17.91 & 16442 &  28 &       \\
 12~56~09.09 & -31~19~34.7 & 17.57 & 17251 &  80 &       \\
 12~56~10.39 & -29~37~19.5 & 18.80 & 15303 &  43 &       \\
 12~56~10.77 & -30~39~08.0 & 18.55 & 15829 &  35 &       \\
 12~56~11.19 & -30~44~07.6 & 18.44 & 15482 &  66 & emiss \\
 12~56~12.12 & -29~52~00.0 & 17.70 & 17081 &  31 &       \\
 12~56~12.35 & -29~02~57.4 & 18.33 & 16169 & 108 &       \\
 12~56~12.37 & -31~18~29.5 & 18.17 & 16905 &  37 &       \\
 12~56~12.92 & -30~08~58.5 & 16.94 & 15336 &  37 &       \\
 12~56~13.54 & -29~37~56.6 & 18.03 & 15261 &  62 &       \\
 12~56~13.64 & -30~36~01.7 & 18.54 & 17341 &  41 &       \\
 12~56~13.79 & -30~44~51.1 & 18.37 & 16728 &  30 &       \\
 12~56~13.79 & -31~06~52.3 & 16.97 & 16579 &  33 &       \\
 12~56~13.97 & -30~23~08.5 & 17.62 &  4684 &  15 & emiss \\
 12~56~15.16 & -30~47~55.4 & 18.29 & 15504 &  54 &       \\
 12~56~16.31 & -30~41~53.5 & 18.03 &  4750 &  15 & emiss \\
 12~56~16.73 & -29~59~04.1 & 18.62 & 25102 &  16 & emiss \\
 12~56~17.74 & -29~09~07.8 & 18.31 & 16736 &  69 &       \\
 12~56~17.94 & -30~38~04.8 & 17.47 & 28270 &  34 &       \\
 12~56~18.12 & -30~03~48.3 & 18.94 & 17424 &  61 &       \\
 12~56~19.14 & -30~57~33.3 & 18.03 & 16670 &  25 &       \\
 12~56~20.22 & -30~40~16.4 & 17.42 & 15327 &  39 &       \\
 12~56~20.64 & -30~52~24.6 & 18.78 & 16973 &  38 &       \\
 12~56~20.73 & -31~11~49.1 & 17.85 & 17611 &  33 &       \\
 12~56~20.78 & -28~59~14.4 & 16.63 & 18041 &  89 &       \\
 12~56~20.80 & -31~14~46.6 & 18.12 & 18190 &  44 &       \\
 12~56~21.28 & -28~50~11.1 & 18.84 & 22115 &  15 & emiss \\
 12~56~21.92 & -28~55~01.8 & 18.28 & 29616 &  73 &       \\
 12~56~22.42 & -31~17~19.9 & 17.88 & 16970 &  62 & emiss \\
 12~56~23.36 & -29~33~46.8 & 18.90 & 15236 &  91 &       \\
 12~56~23.71 & -30~15~25.6 & 18.51 & 15212 &  83 & emiss \\
 12~56~24.15 & -30~40~09.2 & 17.93 & 16904 &  41 &       \\
 12~56~24.80 & -31~25~58.7 & 18.36 & 15776 &  68 &       \\
 12~56~24.86 & -30~05~28.9 & 18.64 & 16948 &  32 &       \\
 12~56~25.33 & -30~13~21.6 & 17.95 & 31929 &  35 & emiss \\
 12~56~27.01 & -29~27~27.1 & 18.98 & 17780 &  59 &       \\
 12~56~27.11 & -30~31~37.4 & 18.58 & 15950 &  31 &       \\
 12~56~27.25 & -29~01~34.5 & 18.25 & 18053 &  36 &       \\
 12~56~27.39 & -31~27~08.2 & 15.86 & 15671 &  30 &       \\
 12~56~27.77 & -29~06~26.6 & 18.67 & 20728 &  15 & emiss \\
 12~56~27.91 & -31~17~08.6 & 16.97 & 17183 &  34 &       \\
 12~56~28.02 & -30~23~39.9 & 18.19 & 16016 &  37 &       \\
 12~56~28.72 & -31~23~26.0 & 18.58 & 17502 &  31 &       \\
 12~56~29.14 & -29~51~21.3 & 18.64 & 15302 &  91 & emiss \\
 12~56~29.94 & -29~53~03.1 & 18.65 & 17189 &  59 &       \\
 12~56~30.25 & -31~17~45.0 & 18.44 & 16691 &  28 &       \\
 12~56~31.73 & -30~38~13.8 & 17.64 & 16980 &  34 &       \\
 12~56~31.99 & -30~17~42.2 & 18.65 & 16057 &  51 &       \\
 12~56~32.25 & -31~24~05.9 & 18.09 & 16383 &  36 &       \\
 12~56~32.59 & -28~57~12.8 & 18.51 & 17728 &  53 &       \\
 12~56~33.39 & -30~30~08.7 & 18.60 & 16390 &  34 &       \\
 12~56~33.58 & -31~03~50.4 & 17.77 & 17451 &  35 &       \\
 12~56~34.15 & -30~07~45.7 & 17.88 & 14889 &  37 &       \\
 12~56~36.54 & -29~55~15.3 & 18.89 & 52587 &  77 &       \\
 12~56~37.99 & -30~27~50.3 & 18.44 & 16536 &  30 &       \\
 12~56~38.34 & -30~47~58.6 & 18.81 & 15201 &  37 &       \\
 12~56~38.78 & -30~10~29.2 & 16.74 & 15965 &  28 &       \\
 12~56~39.09 & -30~51~41.6 & 17.43 & 15534 &  37 &       \\
 12~56~39.29 & -31~01~29.1 & 18.71 & 15382 &  77 &       \\
 12~56~39.88 & -30~07~20.4 & 18.88 & 24592 &  79 &       \\
 12~56~40.76 & -29~52~23.1 & 18.11 & 16643 &  63 &       \\
 12~56~40.85 & -31~03~19.6 & 17.46 & 17175 &  26 &       \\
 12~56~40.93 & -31~20~19.1 & 17.39 & 17200 &  30 &       \\
\noalign{\smallskip}
\hline
\end{tabular}
\end{flushleft}
\end{table}
%-------------------------------------------------------------------------------

%-------------------------------------------------------------------------------
\setcounter{table}{1} 
\begin{table}
\caption[]{ cont. }
\begin{flushleft}
\begin{tabular}{rrrrrr}
\hline\noalign{\smallskip}
$\alpha$ (2000) & $\delta$ (2000) & $b_J$ & $v$ (km/s) & err & notes \\
\noalign{\smallskip}
\hline\noalign{\smallskip}
 12~56~42.13 & -29~45~26.4 & 18.59 & 16970 &  89 & emiss \\
 12~56~42.96 & -30~01~12.7 & 17.75 & 16798 &  33 &       \\
 12~56~43.23 & -30~44~59.3 & 17.36 & 16198 &  32 &       \\
 12~56~43.57 & -30~14~32.4 & 18.07 & 16778 &  28 &       \\
 12~56~44.81 & -30~41~48.8 & 18.53 & 16062 &  41 &       \\
 12~56~44.97 & -31~01~29.8 & 18.18 & 24663 &  40 &       \\
 12~56~46.40 & -30~32~23.1 & 17.30 & 16332 &  39 &       \\
 12~56~47.12 & -29~42~17.7 & 18.12 & 16094 &  69 &       \\
 12~56~48.43 & -30~37~41.2 & 18.40 & 20709 &  35 &       \\
 12~56~48.77 & -28~48~54.2 & 18.79 & 20658 &  15 & emiss \\
 12~56~50.09 & -31~14~05.2 & 18.44 & 18521 &  40 &       \\
 12~56~50.28 & -28~50~47.2 & 17.64 & 15406 &  40 &       \\
 12~56~50.57 & -30~14~25.8 & 17.90 & 17023 &  27 &       \\
 12~56~51.10 & -29~07~44.8 & 18.19 & 72708 &  43 &       \\
 12~56~51.51 & -30~48~54.4 & 18.78 & 16190 &  33 &       \\
 12~56~51.63 & -28~48~44.9 & 18.86 & 20249 &  47 &       \\
 12~56~51.88 & -30~47~40.7 & 17.56 & 16130 &  31 &       \\
 12~56~52.57 & -29~09~55.6 & 18.01 & 24567 &  63 &       \\
 12~56~53.67 & -31~31~43.3 & 17.40 & 15896 &  58 &       \\
 12~56~54.82 & -31~27~15.1 & 16.74 & 16033 &  35 &       \\
 12~56~55.27 & -29~11~03.6 & 18.52 & 16209 & 128 &       \\
 12~56~56.60 & -31~13~47.0 & 18.30 & 26402 &  69 & emiss \\
 12~56~57.39 & -30~30~08.4 & 17.99 & 15145 &  31 &       \\
 12~56~57.40 & -28~56~22.4 & 17.79 & 16681 &  63 &       \\
 12~56~57.41 & -30~24~45.6 & 18.53 & 17235 &  37 &       \\
 12~56~57.86 & -28~59~07.6 & 18.00 & 15562 &  52 &       \\
 12~56~58.09 & -31~19~47.0 & 14.80 & 16881 &  42 &       \\
 12~56~58.36 & -30~06~04.6 & 18.57 & 17162 &  15 & emiss \\
 12~56~58.72 & -31~22~08.8 & 18.19 & 17077 &  39 &       \\
 12~57~00.77 & -29~57~04.8 & 17.33 & 17469 &  41 &       \\
 12~57~02.30 & -30~19~26.3 & 18.37 & 18197 &  33 &       \\
 12~57~02.78 & -29~55~22.6 & 18.93 & 16875 &  15 & emiss \\
 12~57~03.08 & -28~36~29.2 & 18.75 & 20087 &  91 &       \\
 12~57~04.30 & -28~54~24.4 & 18.10 & 17870 &  15 & emiss \\
 12~57~08.63 & -30~01~36.2 & 18.09 & 16617 &  15 & emiss \\
 12~57~08.82 & -31~18~58.4 & 16.97 & 15621 &  32 &       \\
 12~57~09.52 & -29~03~02.7 & 16.93 & 18320 &  55 &       \\
 12~57~11.37 & -31~15~59.4 & 17.55 & 16700 &  31 &       \\
 12~57~11.44 & -30~21~30.2 & 18.50 & 17158 &  45 &       \\
 12~57~11.55 & -28~26~03.5 & 18.68 & 20886 &  39 & emiss \\
 12~57~12.06 & -28~56~45.7 & 17.66 & 15485 &  47 &       \\
 12~57~12.25 & -29~01~55.6 & 18.40 & 21137 &  55 &       \\
 12~57~12.83 & -31~11~12.5 & 18.57 & 15522 &  50 &       \\
 12~57~13.23 & -28~47~37.0 & 17.06 & 20159 &  37 &       \\
 12~57~13.58 & -28~53~49.3 & 18.26 & 24433 &  68 & emiss \\
 12~57~14.45 & -30~20~26.2 & 17.66 & 14347 &  40 &       \\
 12~57~15.22 & -28~22~33.0 & 18.12 & 20750 &  59 &       \\
 12~57~15.94 & -28~45~17.1 & 18.38 & 24585 &  70 & emiss \\
 12~57~16.09 & -29~56~23.4 & 17.42 & 17426 &  65 &       \\
 12~57~16.69 & -29~00~54.9 & 18.33 & 41439 &  79 &       \\
 12~57~16.76 & -31~16~10.3 & 18.04 & 16414 &  36 &       \\
 12~57~17.39 & -30~21~11.9 & 18.20 & 16990 &  24 &       \\
 12~57~17.99 & -30~48~07.9 & 18.80 & 22762 &  80 & emiss \\
 12~57~18.09 & -28~44~17.0 & 16.60 & 20624 &  61 &       \\
 12~57~18.14 & -29~55~12.6 & 17.57 & 17200 &  83 & emiss \\
 12~57~20.22 & -31~11~21.6 & 18.05 & 16575 &  31 &       \\
 12~57~21.30 & -28~30~21.2 & 17.23 & 20096 &  25 &       \\
 12~57~21.84 & -31~23~53.7 & 16.61 & 17815 &  61 &       \\
 12~57~22.00 & -29~53~14.4 & 17.40 &  3110 &  15 & emiss \\
 12~57~23.48 & -30~27~19.2 & 18.54 & 16999 &  35 &       \\
 12~57~23.82 & -30~19~56.9 & 18.10 & 15602 &  33 &       \\
 12~57~24.20 & -28~42~14.8 & 18.28 & 19816 &  29 &       \\
 12~57~24.41 & -28~32~25.6 & 18.37 & 16960 &  28 &       \\
\noalign{\smallskip}
\hline
\end{tabular}
\end{flushleft}
\end{table}
%-------------------------------------------------------------------------------

%-------------------------------------------------------------------------------
\setcounter{table}{1} 
\begin{table}
\caption[]{ cont. }
\begin{flushleft}
\begin{tabular}{rrrrrr}
\hline\noalign{\smallskip}
$\alpha$ (2000) & $\delta$ (2000) & $b_J$ & $v$ (km/s) & err & notes \\
\noalign{\smallskip}
\hline\noalign{\smallskip}
 12~57~24.65 & -29~03~44.0 & 17.24 & 18160 &  62 & emiss \\
 12~57~26.87 & -30~21~28.8 & 18.41 & 16872 &  36 &       \\
 12~57~27.40 & -30~35~52.2 & 17.89 & 16721 &  29 &       \\
 12~57~28.45 & -29~00~20.5 & 16.82 & 18745 &  15 & emiss \\
 12~57~30.65 & -28~38~08.9 & 18.65 & 20310 &  19 & emiss \\
 12~57~31.67 & -31~21~01.5 & 14.84 &  7486 &  40 & emiss \\
 12~57~32.97 & -29~02~30.2 & 18.25 & 22723 &  15 & emiss \\
 12~57~33.65 & -31~21~35.5 & 18.10 & 16623 &  41 &       \\
 12~57~34.88 & -28~18~55.9 & 18.58 & 23010 &  46 & emiss \\
 12~57~34.92 & -29~01~49.3 & 17.25 & 17709 &  70 &       \\
 12~57~36.11 & -28~38~33.4 & 18.60 & 19674 &  43 &       \\
 12~57~39.21 & -28~27~55.2 & 17.21 &  4998 &  95 & emiss \\
 12~57~39.29 & -30~20~40.3 & 18.17 & 16197 &  49 &       \\
 12~57~39.57 & -28~31~21.3 & 18.53 & 20109 &  81 &       \\
 12~57~39.98 & -31~12~13.5 & 18.50 & 22935 &  57 & emiss \\
 12~57~41.08 & -28~24~22.6 & 19.32 & 19749 & 105 & emiss \\
 12~57~43.25 & -28~35~33.8 & 18.65 & 19602 &  32 &       \\
 12~57~43.78 & -28~25~02.7 & 19.27 & 20151 &  95 &       \\
 12~57~44.56 & -28~38~56.2 & 18.80 & 40905 &  76 &       \\
 12~57~46.12 & -31~17~36.5 & 18.45 & 16704 &  33 &       \\
 12~57~46.52 & -28~31~40.9 & 19.01 & 20105 &  96 &       \\
 12~57~46.91 & -28~32~28.1 & 17.61 & 20279 &  32 &       \\
 12~57~47.24 & -28~30~56.2 & 19.18 & 20626 &  89 &       \\
 12~57~47.46 & -28~31~20.1 & 18.58 & 20983 &  56 &       \\
 12~57~48.40 & -28~30~08.7 & 18.49 & 19960 &  43 &       \\
 12~57~48.77 & -28~18~56.3 & 18.78 & 19892 & 145 & emiss \\
 12~57~49.33 & -28~31~20.6 & 16.84 & 20142 &  31 &       \\
 12~57~49.57 & -28~34~55.1 & 17.10 & 17123 &  57 & emiss \\
 12~57~50.05 & -28~28~13.9 & 16.84 & 19970 &  51 &       \\
 12~57~51.18 & -28~40~59.2 & 19.11 & 16977 & 193 & emiss \\
 12~57~51.90 & -28~28~59.2 & 19.35 & 19623 &  97 &       \\
 12~57~52.97 & -28~27~35.6 & 18.21 & 19547 &  53 &       \\
 12~57~53.31 & -28~19~37.8 & 18.39 & 20166 &  44 &       \\
 12~57~53.33 & -30~15~33.3 & 17.96 &  9496 &  90 &       \\
 12~57~53.55 & -28~25~22.6 & 17.19 & 20094 &  91 & emiss \\
 12~57~56.26 & -28~30~21.5 & 18.76 & 14568 & 118 & emiss \\
 12~57~57.06 & -28~19~44.7 & 19.17 & 20562 & 132 &       \\
 12~57~57.20 & -28~16~02.5 & 19.13 & 65710 & 113 &       \\
 12~57~57.52 & -28~20~03.6 & 18.11 & 20382 & 133 & emiss \\
 12~57~57.72 & -30~19~33.1 & 18.49 & 16952 &  84 & emiss \\
 12~57~57.85 & -28~25~23.6 & 17.55 & 17083 &  54 &       \\
 12~57~57.95 & -28~17~20.3 & 19.22 & 23843 &  87 &       \\
 12~58~00.33 & -28~28~15.2 & 18.63 & 17026 &  36 &       \\
 12~58~01.03 & -28~17~52.1 & 19.14 & 19919 &  17 & emiss \\
 12~58~02.10 & -28~25~43.4 & 19.31 & 22527 & 109 & emiss \\
 12~58~03.60 & -28~19~56.5 & 18.60 & 19909 &  51 &       \\
 12~58~04.28 & -28~32~54.2 & 19.35 & 20942 & 115 & emiss \\
 12~58~05.68 & -28~17~03.8 & 17.93 & 19736 &  22 & emiss \\
 12~58~06.16 & -28~17~50.4 & 18.48 & 19674 &  62 &       \\
 12~58~07.12 & -30~28~11.2 & 17.61 & 22754 &  12 & emiss \\
 12~58~09.40 & -28~28~25.4 & 19.40 & 20370 &  83 &       \\
 12~58~13.71 & -28~29~21.9 & 19.22 & 20396 & 200 & emiss \\
 12~58~14.02 & -28~38~09.7 & 19.08 & 40658 & 122 & emiss \\
 12~58~15.50 & -28~16~20.4 & 17.84 & 23416 &  61 &       \\
 12~58~19.32 & -28~32~33.2 & 18.73 & 20094 & 104 & emiss \\
 12~58~20.66 & -28~30~13.4 & 18.32 & 20731 &  43 &       \\
 12~58~20.92 & -28~36~15.1 & 19.22 & 20208 &  98 & emiss \\
 12~58~23.24 & -28~14~50.4 & 18.90 & 22484 & 123 & emiss \\
 12~58~24.08 & -28~29~29.5 & 17.72 & 20444 &  36 &       \\
 12~58~24.89 & -28~40~54.6 & 18.46 & 19183 &  86 & emiss \\
 12~58~26.56 & -28~16~45.1 & 18.38 & 23148 &  68 &       \\
 12~58~26.71 & -28~29~43.8 & 17.05 & 20719 &  39 &       \\
 12~58~29.00 & -28~37~13.8 & 18.28 & 20379 &  48 &       \\
\noalign{\smallskip}
\hline
\end{tabular}
\end{flushleft}
\end{table}
%-------------------------------------------------------------------------------

%-------------------------------------------------------------------------------
\setcounter{table}{1} 
\begin{table}
\caption[]{ cont. }
\begin{flushleft}
\begin{tabular}{rrrrrr}
\hline\noalign{\smallskip}
$\alpha$ (2000) & $\delta$ (2000) & $b_J$ & $v$ (km/s) & err & notes \\
\noalign{\smallskip}
\hline\noalign{\smallskip}
 12~58~31.92 & -28~32~53.8 & 18.61 & 40941 &  81 & emiss \\
 12~58~36.56 & -28~29~42.5 & 19.09 & 32533 &  47 & emiss \\
 12~58~37.94 & -28~41~33.7 & 18.40 & 19344 &  44 & emiss \\
 12~58~39.65 & -28~37~27.7 & 17.05 & 21062 &  57 &       \\
 12~58~39.91 & -28~15~27.9 & 17.93 & 19528 & 116 &       \\
 12~58~40.52 & -28~28~45.8 & 17.58 & 20928 &  13 & emiss \\
 12~58~41.30 & -28~41~18.8 & 19.04 & 31627 & 201 &       \\
 12~58~42.49 & -28~32~54.6 & 19.17 & 41215 & 165 & emiss \\
 12~58~43.41 & -28~33~10.0 & 17.45 & 20823 &  71 &       \\
 12~58~43.64 & -28~24~06.3 & 19.06 & 20619 &  57 & emiss \\
 12~58~55.02 & -28~38~15.4 & 18.59 & 20706 & 155 & emiss \\
 12~58~57.67 & -28~23~54.5 & 19.30 & 60140 &  49 & emiss \\
 12~59~01.74 & -28~33~15.9 & 19.25 & 15830 &  79 & emiss \\
 12~59~12.39 & -28~23~58.5 & 19.15 & 60356 &  35 &       \\
\noalign{\smallskip}
\hline
\end{tabular}
\end{flushleft}
\end{table}
%-------------------------------------------------------------------------------

\end{document}